\documentclass[%
 reprint,
nofootinbib,
 amsmath,amssymb,
 aps,
prx,
]{revtex4-2}
\usepackage[english]{babel}
\usepackage{graphicx}
\usepackage{dcolumn}
\usepackage{bm}
\usepackage{lmodern}                       
\usepackage[T1]{fontenc}                   
\usepackage{ae}   
\usepackage[utf8]{inputenc}    
\usepackage{xfrac}  
\usepackage{physics}
\RequirePackage{dsfont}                      

\usepackage[dvipsnames,x11names]{xcolor} 	
\usepackage{comment}    
\usepackage{units}
\usepackage[normalem]{ulem}
\usepackage{blindtext}  
\usepackage{float}  
\usepackage{bbold} 
\usepackage[colorlinks=true,citecolor=OliveGreen,linkcolor=Purple,urlcolor=Magenta]{hyperref} 	
\usepackage{soul}
\newcommand{\map}[1]{\mathcal{#1}}
\newcommand{\dket}[1]{\vert#1\rangle\hspace{-.8mm}\rangle}

\newcommand{\dketbra}[2]{\vert #1 \rangle \hspace{-.8mm} \rangle \hspace{-.4mm} \langle\hspace{-.8mm}\langle #2 \vert}

\newcommand{\id}{\mathds{1}}
\renewcommand{\H}{\mathcal{H}}
\renewcommand{\L}{\mathcal{L}}

\newcommand{\pink}[1]{\textcolor{magenta}{#1}}
\begin{document}
\preprint{APS/123-QED}
\title{Higher-order Process Matrix Tomography of a passively-stable Quantum SWITCH}

\author{Michael Antesberger$^{1,\dagger}$}
\author{Marco Túlio Quintino$^{2}$}
\author{Philip Walther$^{1,4}$}
\author{Lee A. Rozema$^1$}

\affiliation{
$^1$University of Vienna, Faculty of Physics, Vienna Center for Quantum Science and Technology (VCQ) {\&} Research Network Quantum Aspects of Space Time (TURIS) \\
$^2$ Sorbonne Université, CNRS, LIP6, F-75005 Paris, France \\
$^3$ Institute for Quantum Optics and Quantum Information (IQOQI), Austrian Academy of Sciences, Boltzmanngasse 3, 1090 Vienna, Austria \\
$^4$ Christian Doppler Laboratory for Photonic Quantum Computer, Faculty of Physics,  University of Vienna, 1090 Vienna, Austria\\
$^\dagger$Correspondence to:  michael.antesberger@univie.ac.at.\\
}


\date{\today}

\begin{abstract}  
The field of indefinite causal order (ICO) has seen a recent surge in interest.
Much of this research has focused on the quantum SWITCH, wherein multiple parties act in a superposition of different orders in a manner transcending the quantum circuit model. 
This results in a new resource for quantum protocols, and is exciting for its relation to issues in foundational physics.
The quantum SWITCH is also an example of a higher-order quantum operation, in that it not only transforms quantum states, but also other quantum operations.
To date, no higher-order quantum operation has been completely experimentally characterized.
Indeed, past work on the quantum SWITCH has confirmed its ICO by measuring causal witnesses or demonstrating resource advantages, but the complete process matrix has only been described theoretically. 
Here, we perform higher-order quantum process tomography. 
{However, doing so requires exponentially many measurements with a scaling worse than standard process tomography.
We overcome this challenge by creating a new} \emph{passively-stable} fiber-based quantum SWITCH using {active optical elements to deterministically generate and manipulate} time-bin encoded qubits.
{Moreover, our new architecture for the quantum SWITCH can be readily scaled to multiple parties.}
By reconstructing the process matrix, we estimate its fidelity and tailor different causal witnesses directly for our experiment. To achieve this, we measure a set of tomographically complete settings, that also spans the input operation space. 
Our tomography protocol allows for the characterization and debugging of higher-order quantum operations with and without an ICO, while our experimental time-bin techniques could enable the creation of a new realm of higher-order quantum operations with an ICO.
\end{abstract}

\pacs{Valid PACS appear here}

\maketitle

\vspace{-8mm}
\section{Introduction}
The formalism of higher-order quantum operations (HOQO) provides a framework to view quantum operations as objects that can be subjected to transformations \cite{selinger_valiron_2009,chiribella08_arquitechture,gutoski07,chiribella09Networks,bisio19theoretical,oreshkov2012quantum}. 
This framework is particularly useful for analysing causality in quantum mechanics.
Since it was first realized that quantum mechanics allows for processes with an indefinite causal order (ICO) \cite{hardy2007towards,Chiribella2013,oreshkov2012quantum}, the field of quantum causality has seen an increasing level of interest \cite{brukner2014quantum}.
These processes are interesting for a variety of foundational topics \cite{baumeler2014perfect,giarmatzi15_causallySEP,baumeler2016space,Arajo2015,Branciard2016,zych2019bell}, and also because it has been recognized that they can lead to ICO-based enhancements that go even beyond ``normal quantum technology.'' 
Examples of these advantages include applications in quantum computing \cite{hardy2009quantum,Araujo2014,Araujo2017,renner22advantage}, quantum communication \cite{feix2015,Guerin2016,Ebler2018,jia2019causal,kristjansson2019,chiribella2019quantum,chiribella2021indefinite,caleffi2020,procopio2020sending}, channel discrimination\cite{ChiribellaSWITCH,bavaresco21a}, metrology \cite{zhao2019}, reversing quantum dynamics~\cite{quintino19_inversion,Schiansky22}, and even thermodynamics \cite{felce2020}.
Experimentally, work has focused on either implementing various protocols  \cite{Lorenzo,goswami2018communicating,wei2019experimental,guo2020experimental,RubinoCommunication2021,taddei2021computational,pang2023experimental,liu2023experimentally} or verifying the ICO of a given experimental implementation \cite{rubino2017,goswami2018,Rubino2022experimental,cao2022experimental}.

In spite of this large body of work, there has not yet been a complete experimental characterization of process with an ICO.
Instead, previous work on ICO has mainly focused on designing and measuring witnesses to essentially provide a yes or no answer to the question ``does this process have an indefinite causal order?''
On the one hand, this is because no concrete protocol for a complete characterization has yet been presented. On the other hand, this is because the number of experimental settings required for a complete characterization has been prohibitive in past experiments.
Here, we overcome both of these hurdles, first presenting a protocol to perform ``higher-order process matrix tomography,'' and then implementing a new experimental method to realize the quantum SWITCH based on time qubits, based on the proposal of \cite{rambo2016functional}.
Our new passively-stable implementation is based on active optical elements, and it allows us to acquire sufficient data (estimating almost 10,000 distinct probabilities) to fully reconstruct a process matrix demonstrating an ICO for the first time.

\begin{figure}
    \centering\includegraphics[scale=0.1]{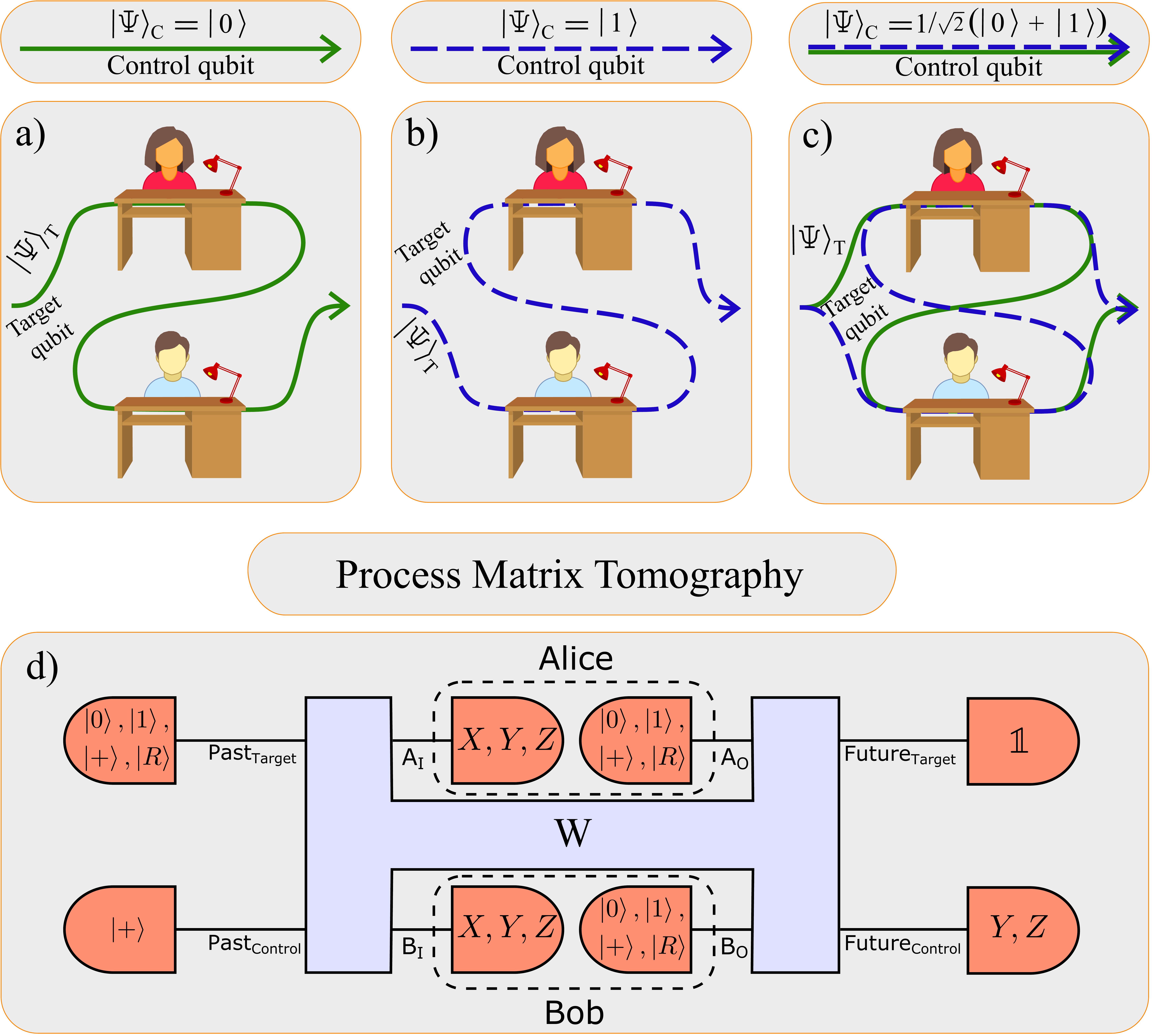}
    \caption{\textbf{The quantum SWITCH.} Panel \textbf{a)} (\textbf{b)}) shows the causally ordered process with the control in state $\ket{0}_C$ ($\ket{1}_C$), where Alice (Bob) acts before Bob (Alice) on the target system. \textbf{c)} A superposition of orders with the control qubit in the state $\frac{1}{\sqrt{2}}\ket{0}_C + \ket{1}_C$. \textbf{d)} The principle of the process tomography on the quantum SWITCH. Alice and Bob perform projective measurements in three different basis and then prepare four different linearly independent states in their output. The same input states are prepared at the past-target. The past-control is fixed to the superposition state for generating the indefinite order. Finally, the future-control is measured in different basis, while the future-target is traced out (the two measurements shown are those that are implemented experimentally; to span the space, a third measurement is needed).
    }
    \label{fig::cartoon}
\end{figure}

In the two-party quantum SWITCH, Alice and Bob each act on a target system (typically taken to be a qubit) in their local laboratories.
This target qubit is sent first to one party and then to the other.
The order in which the target qubit is shared between the two is coherently controlled via a second \emph{control} qubit. 
If the control qubit is prepared in a superposition state, then the two parties act on the target qubit in a superposition of orders (Fig. \ref{fig::cartoon}a-c). 
The quantum SWITCH and the so-called quantum time flip \cite{chiribella2022quantum,guo2022experimental,stromberg2022experimental}, are the only processes which do not respect our standard notions of causality that have been experimentally implemented to date.
The quantum SWITCH is an example of a HOQO, in the sense that its inputs are not only the control and target qubits, but also Alice and Bob's operations.

All experimental realizations of the quantum SWITCH have been accomplished by encoding both the control and the target systems in a single photon.
Typically, the control system is encoded in a path degree-of-freedom, which then determines the order that the photon is routed between the two parties. 
In practice, this means that a photon is placed in a superposition of the two paths using a beamsplitter, and these paths are then looped between two parties in a manner mimicking the paths of Fig. \ref{fig::cartoon}c.  The parties then act on a different degree of freedom, such as polarization \cite{Lorenzo}, time bins \cite{wei2019experimental}, or orbital angular momentum \cite{goswami2018}.
The result of all of these approaches is essentially a Mach-Zehnder interferometer, which must be phase stabilized.%
\footnote{Polarization has also been used \cite{goswami2018,goswami2018communicating} as a control system, but also in this case the polarization is used to route the photon into two paths in superposition, which also requires a stable phase.}
Stabilizing the phase for long enough to acquire the required data for full higher-order quantum process tomography presents a daunting experimental challenge.

We overcome this challenge by implementing a new passively-stable quantum SWITCH. In our experiment, the control system is encoded in a time-bin qubit, the target qubit is encoded in the polarization of the same photon, and active optical switches are used to route the photon between the two parties in superposition of both orders, as in the theoretical proposal of \cite{rambo2016functional}.
While two other experiments have achieved an intrinsically stable phase, using a Sagnac-like approach \cite{wei2019experimental,stromberg2022demonstration}, it is not clear how to scale these methods to multiple-parties.
Our approach, however, has straight-forward generalization to multiple parties \cite{rambo2016functional}, making it a promising new experimental method to create an ICO.

Recently, there has been some discussion in the community whether such photonic implementations of the quantum switch are simulations of an ICO \cite{paunkovic2020causal,vilasini2022embedding}, with some concluding that they may only have an ICO in a ``weak sense'' \cite{ormrod2022causal}, while others conclude that the experiments do have an ICO \cite{oreshkov2019time, DeLaHemette_quantumDiffeomorphisms}, or that they at least have a quantifiable resource advantage \cite{fellous2022comparing}.
Here we do not address this debate, but we make use of the mathematical formalism of processes matrices and HOQO to describe our physical experiment. 

One method to certify ICO is made via the violation of a so-called causal inequality \cite{oreshkov2012quantum}. 
This is a device-independent technique, similar to the use of a Bell violation to verify entanglement \cite{guhne2009entanglement}.
Unfortunately, it is not yet known if one can implement a quantum process that deterministically violates a causal inequality; moreover, it has been shown that the quantum SWITCH cannot violate causal inequalities \cite{Araujo2014}.
Instead, in the first implementation of the quantum SWITCH \cite{Lorenzo}, the ICO was indirectly proven by performing a game, where a player has to decide if two unitary gates either commute or anti-commute (see App.~\ref{sec::game}).
By winning that game more than one could with a definite causal order, it was concluded the experiment did not have a definite causal order.
This method can, in fact, be reframed, in terms of a causal witness \cite{Arajo2015}.
A causal witness is a measurement which can be used to verify if a process is causally non-separable (i.e. if it has an ICO), and it has been experimentally implemented for the quantum SWITCH \cite{Rubino2022experimental,goswami2018}.
Unlike a causal inequality, a causal witness is not device independent, requiring the assumption that the experimenter knows the correct quantum description of the experiment.
Recently, progress has been made by relaxing the complete device independent approach, allowing the certification of causal non-separability under semi-device independent assumptions \cite{bavaresco19,cao2022experimental,Dourdent02}, Bell locality-like assumptions \cite{zych2019bell,Rubino2022experimental}, and additional device independent no-signalling assumptions \cite{derLugt2022,Gogioso06}.

In this work, we implement full experimental higher-order process tomography of the quantum SWITCH. For this goal, we generalise the ideas from quantum state and quantum process tomography~\cite{chiribella08_arquitechture,ziman08PPOVM,chiribella09Networks,chuang1997prescription,poyatos1997complete,o2004quantum,bisio17tomo,mohseni2008quantum,rozema2014optimizing} to tackle tomography of arbitrary higher-order processes,  including those without a definite causal order. In particular, we show that it is possible to construct tomographically complete measurement settings on arbitrary quantum processes by using a tomographically complete set of input states, spanning the input state space; a tomographically complete set of measurement-repreparation channels, spanning the input operation spaces; and a tomographically complete set of quantum measurements spanning the output state space (see Fig. \ref{fig::cartoon}d). 
We then employ these ideas to experimentally perform higher-order quantum process tomography on the quantum SWITCH.

The rest of the paper is organized as follows. In Section~II we introduce the theory of quantum process matrices, using the quantum SWITCH as a paradigmatic example, and we present our causal tomography protocol.  In Section III, we discuss our new passively-stable architecture for the quantum SWITCH. Section IV presents our experimental results, and we conclude in Section V.

\section{Theory}
\label{sec:theory}
\subsection{Process matrices and the quantum switch}

The expression quantum process is a general term used to refer to the dynamics of quantum systems, and its precise meaning may depend on the context. For instance, when analysing transformations between quantum states, quantum process refers to a quantum channel, which may be unitary (associated with closed quantum systems and mathematically described by unitary operators) or non-unitary (associated with open quantum systems and mathematically described by Completely Positive Trace Preserving (CPTP) linear maps). In this scenario of transformations between states, one can experimentally determine the dynamics by means of what is known as quantum process tomography \cite{chuang1997prescription,poyatos1997complete,o2004quantum}. To do so, a complete set of known quantum states is fed into an unknown quantum process $\mathcal{E}$ and a complete set of measurements is performed on the output of the underlying process for each input state \cite{rozema2014optimizing}. 
When performing standard process tomography, one reconstructs, for example, the chi matrix $\chi$, which takes quantum sates as inputs and returns quantum states as outputs \cite{NielsenChuang}. The chi matrix is often called a process matrix; however, we stress that this chi or process matrix is different from the process matrices discussed in the field of HOQO and ICO, the case which we address here.

In this work, we analyse transformations between quantum channels, hence we use the word process to describe the dynamics between quantum operations. 
An operation that transforms a quantum operation is sometimes referred to as a higher-order quantum operation. Such operations have found applications in several branches of quantum information processing \cite{gutoski07,chiribella08_arquitechture,chiribella09Networks,bisio10,sedlak18,quintino19_inversion,bavaresco21b}. The formalism of higher-order transformations, is also particularly useful for describing quantum processes which may not respect a definite causal order, such as the notorious quantum SWITCH~\cite{Chiribella2013,ChiribellaSWITCH}, the main object analysed in this work. In its most general form, the quantum SWITCH is a process that, transforms a pair of unitary operators $(U_A,U_B)$ into another operator which is a coherent superposition of the composition $U_A U_B$ and $U_BU_A$. In mathematical terms, the quantum SWITCH is the transformation
\begin{align} \label{eq:switch}
    (U_A,U_B) \mapsto \ketbra{0}{0} \otimes U_AU_B + \ketbra{1}{1} \otimes U_BU_A, 
\end{align}
where the first system of the RHS of Eq. \ref{eq:switch} is referred to as the control system, since the order in which $U_A$ and $U_B$ will be performed may be controlled by setting the control qubit state. The second system is referred to the target system, since it is the system on which the unitary operators act.

\subsection{Choi-Jamiołkowski isomorphism and quantum operations} \label{sec:Choi}
Higher-order transformations such as the quantum SWITCH may be conveniently described by means of a \textit{process matrix}~\cite{oreshkov2012quantum}, a formalism which is heavily based on the  Choi-Jamiołkowski (CJ) isomorphism~\cite{jamiolkowski1972linear,choi1975completely}, a method to represent linear maps as linear operators, and linear operators as vectors. Let $\H_\text{in}$ and $\H_\text{out}$ be finite linear (Hilbert) spaces associated with the input and output. Let $U:\mathcal{H}_{\text{in}}\to\mathcal{H}_{\text{out}}$ be a linear operator, its process vector $\dket{U}\in\mathcal{H}_{\text{in}} \otimes \mathcal{H}_{\text{out}}$  is the defined as\footnote{Upper-indices on vectors and operators will be used to indicate where these objects act. For instance, in $\dket{U}^{\mathcal{H}_{\text{in}}\mathcal{H}_{\text{out}}}$, the upper indices indicate that  $\dket{U}\in\mathcal{H}_{\text{in}} \otimes \mathcal{H}_{\text{out}}$. Whenever clear from context, we might drop the upper indices to avoid an overly heavy notation.}
\begin{equation}    
\dket{U}^{\mathcal{H}_{\text{in}}\mathcal{H}_{\text{out}}}:= \sum_i \ket{i}
\otimes U\ket{i},
\label{eq::U}
\end{equation}
where $\{\ket{i}\}_i$ is the computational basis.  Notice that, the Choi vector of the identity operator is given by
\begin{align}
    \dket{\id}=\sum_{i} \ket{i}\otimes\ket{i},
\end{align}
which is equivalent to a maximally entangled state up to normalisation.

Let $\L(\H)$ be the set of all linear operators acting on $\H$. 
Let $\mathcal{C}:\L(\H_\text{in})\to\L(\H_\text{out})$ be a linear map, its Choi operator $C\in\L(\H_\text{in}\otimes\H_\text{out})$ is defined as
\begin{equation}
    C^{\mathcal{H}_{\text{in}}\H_\text{out}}:=\sum_{ij} \ketbra{i}{j}\otimes \mathcal{C}(\ketbra{i}{j}).
\end{equation}
Then, the action of any linear map $\mathcal{C}$ on a state $\rho$ can be written in terms of the Choi operator $C$ as
\begin{align}
    \map{C}(\rho)=\Tr_\text{in}\left({\rho^T}^{\H_\text{in}}\otimes\id^{\H_\text{out}} \, C^{\H_\text{in}\H_\text{out}}\right),
\end{align}
where $\rho^T$ is the transpose of $\rho$ in the computational basis and $\rho$ is an arbitrary density operator acting on $\H_\text{in}$.

The Choi-Jamiołkowski isomorphism is very useful to represent quantum operations, due to the fact that a linear map $\mathcal{C}:\L(\H_\text{in})\to\L(\H_\text{out})$ is completely positive if and only if its Choi operator $C\in\L(\H_\text{in}\otimes\H_\text{out})$ respects $C\geq0$, and the map $\mathcal{C}$ is trace-preserving if and only if $\Tr_\text{out}(C)=\id^{\H_\text{in}}$. Since quantum channels are completely positive trace-preserving maps, all quantum channels have a simple and direct characterisation in terms of their Choi operators.
Before finishing this subsection we also remark that if $\mathcal{C}$ is a unitary channel, that is $\mathcal{C}(\rho)=U\rho U^\dagger$ for some unitary operator $U$, direct calculation shows that its Choi operator may be written as $C=\dketbra{U}{U}$, where $\dket{U}$ is defined in Eq. \ref{eq::U}.

A quantum instrument is a quantum operation which has a classical and a quantum output, and it formalises the concept of a quantum measurement which has a post-measurement quantum state. Mathematically, a set of linear maps $\{\mathcal{C}_i\}_i$,  $\mathcal{C}_i:\L(\H_\text{in})\to\L(\H_\text{out})$ is a quantum instrument if all $\mathcal{C}_i$ are completely positive and $\mathcal{C}:=~\sum_i \mathcal{C}_i$ is trace preserving. In the Choi operator picture, this is equivalent to having $C_i\geq0$ and $\Tr_\text{out}\left(\sum_i C_i \right)=\id^{\H_\text{in}}$. A simple and useful class of quantum instrument is the class of \textit{measure-and-reprepare} instruments. In its most basic form, a measure and reprepare instrument simply performs a measurement described by the operators\footnote{A set of operators $\{M_i\}$ represents a quantum measurement (that is, it is a POVM) if $M_i\geq0$ and $\sum_i M_i=\id$.} $\{M_i\}$, and reprepares some fixed state $\sigma$. Its linear map is described by $\mathcal{R}_i(\rho):=\Tr(\rho\, M_i) \sigma$, and its Choi operators are given by $R_i\in\L\left(\H_\text{in}\otimes\H_\text{out}\right)$, with $R_i= M_i^T\otimes \sigma$.

\subsection{The quantum switch as a process matrix}
We are now in position to present process matrices which describe transformations between the quantum channels of different parties. We start by presenting the process vector describing Fig~\ref{fig::cartoon}a, which is simply a process where a system flows freely from a common past target space $\H_{P_\text{t}}$ to Alice's input space $\H_{A_\text{in}}$.  
Alice may perform an arbitrary operation as the state goes from $\H_{A_\text{in}}$ to $\H_{A_\text{out}}$. Later, the state goes freely from Alice's output space $\H_{A_\text{out}}$ to Bob's input space $\H_{B_\text{in}}$. Bob may then perform an arbitrary operation as the state goes from $\H_{B_\text{in}}$ to $\H_{B_\text{out}}$. Finally, the state goes freely from Bob's output space $\H_{B_\text{out}}$ to a common future target space $\H_{F_\text{t}}$.
The process vector of this quantum process is
\begin{equation}
\ket{A\to B} := \dket{\id}^{P_\text{t}, {A}_{\text{in}}} \otimes \dket{\id}^{{A}_{\text{out}}, {B}_{\text{in}}} \otimes \dket{\id}^{{B}_{\text{out}},F_\text{t}} \,,   
\label{eq::AB vector}
\end{equation}
and its process matrix is given by
\begin{equation}
    W_{A \to B}:=\ketbra{A \to B}{A \to B},
\end{equation}
where $A \to B$ indicates that Alice acts before Bob.
{(Note that we have not included Alice or Bob's operations in this description; this will be introduced in Sec. \ref{sec:meas}.)}
Analogously, we may define the process where Bob acts before Alice, which will lead to a process vector
\begin{equation}
    \ket{B \to A} := \dket{\id}^{\mathcal{H}_{\text{P}}, \mathcal{B}_{\text{in}}} \otimes \dket{\id}^{\mathcal{B}_{\text{out}}, \mathcal{A}_{\text{in}}} \otimes \dket{\id}^{\mathcal{A}_{\text{out}}, F_{\text{t}}} \,,
\label{eq::BA vector}
\end{equation}
and its process matrix is given by
\begin{equation}
    W_{B \to A}:=\ketbra{B \to A}{B \to A}.
\end{equation}

The quantum SWITCH is a process which allows one to coherently alternate between $ \ket{A \to B} $ and $ \ket{B \to A}$. For that, we allow the common past and common future to have another system, denoted as a control system, which will be able to coherently alternate between the ordered process. More formally, the common past and common future space are now described by $\H_P=\H_{P_\text{c}}\otimes~\H_{P_\text{t}}$ and  $\H_F=\H_{F_\text{c}}\otimes\H_{F_\text{t}}$  respectively, and the Choi vector of the quantum switch is given by 
\begin{align}
    \ket{w_\text{switch}} := \ket{0}^{{P}_c}&\otimes \ket{A \to B} \otimes \ket{0}^{{F}_c} \nonumber \\
    &+ \ket{1}^{{P}_c}\otimes \ket{B \to A} \otimes \ket{1}^{{F}_c}, 
\end{align}
which corresponds to the process matrix
\begin{equation} \label{eq:switch_full}
    W_\text{switch}:=\ketbra{w_\text{switch}}{w_\text{switch}}
\end{equation}

Almost all known applications of the quantum SWITCH, \textit{e.g.}, computational advantages~\cite{Araujo2014}, channel discrimination~\cite{Chiribella2013}, reducing communication complexity~\cite{Guerin2016}, semi-device-independent~\cite{bavaresco19} and device-independent certification of indefinite causality~\cite{derLugt2022,gogioso22}, do not require the general form of the quantum SWITCH is presented in Eq.~\eqref{eq:switch_full}. Rather, in such applications, one starts with the control qubit in the $\ket{+}:=\frac{\ket{0}+\ket{1}}{\sqrt{2}}$, so that the process state corresponds to a coherent superposition of processes described by:
\begin{align}
    \ket{w_\text{switch}^+} := \frac{\ket{A \to B} \otimes \ket{0}^{{F}_c} + \ket{B \to A} \otimes \ket{1}^{{F}_c}}{\sqrt{2}}.
\end{align}
Additionally, for all such applications, one does not make use of the future target system, hence this qubit is often discarded. Mathematically, discarding a system corresponds to the partial trace. Hence, we construct the simplified version of the SWITCH as
\begin{equation} \label{eq:switch_s}
    W_\text{s}^+:=\Tr_{F_\text{t}} \left(\ketbra{w_\text{switch}^+}{w_\text{switch}^+}\right)
\end{equation}

In this work, we focus on the simplified quantum SWITCH, and as is usual in the literature, we use  ``quantum SWITCH'' to refer to the process described in Eq.~\eqref{eq:switch_s}.

\subsection{The general process matrix formalism}\label{subsec:general_process}
The process matrix formalism allows one to assign a matrix which perfectly describes transformations between arbitrary quantum objects, in particular, to transform quantum channels into quantum channels. The normalisation constraints from quantum channels (or more general quantum objects) and a generalised notion of completely positive inputs lead to constraints on valid process matrices. In a nutshell, when focused on quantum channels, a matrix $W$ is a process matrix if it is positive semidefinite and respect a set of affine constraints arising from the channel normalisation conditions. These affine constraints are described in several Refs. such as \cite{oreshkov2012quantum,Araujo2014} and may be viewed as causal constraints (for instance, they prevent local loops or the possibility of obtaining negative probabilities via Born's rule).

\subsection{Measuring a process matrix}
\label{sec:meas}
One of the main applications of the process matrix formalism is to provide mathematical methods to analyse the dynamics of a quantum process and to predict the outcomes of measurements performed on a quantum process. Below, we describe the scenario considered in this work:

\begin{figure}
    \centering
    \includegraphics[width=\linewidth]{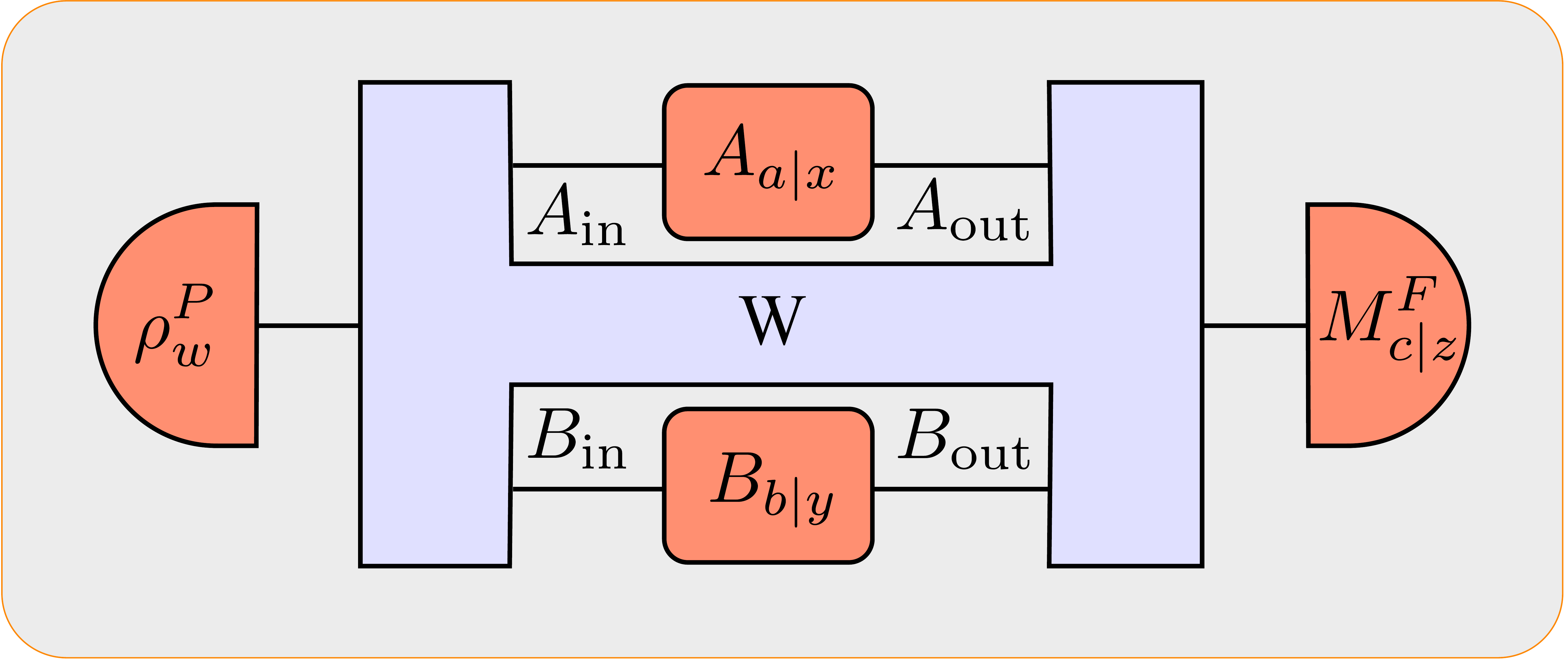}
    \caption{\textbf{Probing a quantum process $W$.} Pictorial illustration on how to probe a quantum process bipartite quantum process $W$ with a common past and common future. Here $\rho_{w}$ are quantum states, $\{A_{a|x}\}$ and $\{B_{b|y}\}$ are quantum instruments, and $\{M_{c|z}\}$ are quantum measurements.  }
    \label{fig:probing}
\end{figure}
\begin{enumerate}
    \item $W\in\L\left(\H_P \otimes \H_{A_\text{in}}\otimes\H_{A_\text{out}}\otimes\H_{B_\text{in}}\otimes\H_{B_\text{out}}\otimes\H_{F}\right)$ is the process matrix which describes a bipartite scenario with a common past {(target)} and common  future (control).
    \item $\rho\in\L(\H_{P})$ is a quantum state on the common past {(target)} space.
    \item $A_a\in\L(\H_{A_\text{in}}\otimes\H_{A_\text{out}})$ are the Choi operators of an instrument on Alice's space.
    \item $B_b\in\L(\H_{B_\text{in}}\otimes\H_{B_\text{out}})$ are the Choi operators of an instrument on Bob's space.
    \item $M_c\in\L(\H_F)$ are the measurement operators on the common future (control) space.
\end{enumerate}
In the scenario described above, if $W$ is the process matrix, one inputs the state $\rho$ into the common past, Alice performs the instrument $\{A_a\}$, Bob performs the instrument $\{B_b\}$, the measurement $\{M_c\}$ is performed in the future, the probability that Alice obtains the outcome $a$, Bob obtains the outcome $b$, and the future obtains the outcome $c$ is given by
\begin{align}
    p(a,b,c)=\Tr\left(W \, \left(\rho^{P}\otimes A_a^{A_\text{in}A_\text{out}}\otimes B_b^{B_\text{in}B_\text{out}}\right)^T\otimes M_c^{F} \right).
\end{align}

In practice, it is often convenient to have indices to label states, instruments, and measurements. In this work, we will then use $\{\rho_w\}$ to denote a set of states acting in the common past, $\{A_{a|x}\}$ for a set of instruments in Alice's space ($a$ labels the classical outcome of the instrument and $x$ the choice of instrument), $\{B_{b|y}\}$ for a set of instruments in Bob's space ($b$ labels the classical outcome of the instrument and $y$ the choice of instrument), and $\{M_{c|z}\}$ for a set of measurements in the future space ($c$~labels the classical outcome and $z$ the choice of measurement). We can then define the \emph{setting operators}\footnote{We use the phrase ``setting operator'', since each of these operators will correspond to a single experimental setting.} as
\begin{align}
   S_{abc|xyzw}:=
   \left( \rho^{P}_w\otimes A_{a|x}^{A_\text{in}A_\text{out}}\otimes B_{b|y}^{B_\text{in}B_\text{out}}\right)^T\otimes M_{c|z}^{F} 
\end{align}
which leads us to the so-called ``generalised Born's rule'':
\begin{align}
    p\left(abc|xyzw\right)=\Tr\left(W \,S_{abc|xyzw}\right).
\end{align}

\subsection{Process matrix tomography}
The goal of quantum tomography is to completely characterise a quantum object by performing known measurements on it. 
Before discussing process matrix tomography, we revisit the standard case of quantum state tomography, where one aims to characterise an unknown state by analysing the outcomes obtained after performing known measurements on it. If $M_{a|x}\in\L(\mathbb{C}_d)$ are known measurement operators, one can make use of the probabilities $p(a|x)=\Tr(\rho\, M_{a|x})$ to uniquely reconstruct the unknown state $\rho$. When the set of operators $\{M_{a|x}\}$ spans the linear space of $\L(\mathbb{C}_d)$, the operator $\rho$ may be obtained via $p(a|x)$ by standard linear inversion methods.

For qubit states, a standard set of tomographically complete measurements is formed by the three Pauli observables $X$, $Y$, and $Z$, which are associated with the measurement operators via their eigenprojectors: $\big\{\ketbra{+}{+},\ketbra{-}{-}\big\}$, $\big\{\ketbra{y_+}{y_+},\ketbra{y_-}{y_-}\big\}$, $\big\{\ketbra{0}{0},\ketbra{1}{1}\big\}$, respectively, where $\ket{\pm}=\frac{\ket{0}\pm \ket{1}}{\sqrt{2}}$ and $\ket{y_\pm}=\frac{\ket{0}\pm i\ket{1}}{\sqrt{2}}$. In particular, the standard measurement operators from the set
\begin{align}
\label{eq:stateSet}
    \mathcal{S}:=\{\ketbra{\psi_i}{\psi_i}\}_{i=1}^4,
\end{align} where
\begin{align}
  \label{eq:states1}
  &\ketbra{\psi_1}{\psi_1}:=\ketbra{0}{0}, \hspace{1.8mm}  \ketbra{\psi_2}{\psi_2}:=\ketbra{1}{1},  \\
   &\ketbra{\psi_3}{\psi_3}:=\ketbra{+}{+},  \ketbra{\psi_4}{\psi_4}:=\ketbra{y_+}{y_+}.
\label{eq:states2}
\end{align}
These measurements are linearly independent, forming a (non-orthonormal) basis for $\L(\mathbb{C}_2)$.

We now consider the task of performing tomography of a {qubit} channel. As discussed in section \ref{sec:Choi}, every quantum channel $\mathcal{C}:\L(\H_\text{in})\to\L(\H_\text{out})$, can be represented by its Choi operator $C\in\L(\H_\text{in}\otimes \H_\text{out})$. 
In this case, tomography can be carried out by preparing a set of states $\{\rho_w\}_w$, $\rho_w\in\L(\H_\text{in})$ and performing a complete set of measurements on each state.
For qubits, the standard measurements are
\begin{align}
\label{eq:measSet}
    \mathcal{M}:=\{M_{i|j}\}_{i=1,j=1}^{i=2,j=3},
\end{align} {where $M_{i|j}$ is a POVM element, the label $i$ stands for the outcomes, and $j$ for the choice of measurements. Hence, $\mathcal{M}$ is a set with three dichotomic measurements with POVM elements given by}
\begin{align}
\label{eq:measDef}
  &M_{1|1}:= \ketbra{0}{0}, \hspace{4mm} M_{2|1}:=\ketbra{1}{1} \\
  \label{eq:measDef2}
  &M_{1|2}:= \ketbra{+}{+}, \hspace{2.3mm} M_{2|2}:=\ketbra{-}{-} \\
   &M_{1|3}:= \ketbra{y_+}{y_+}, M_{2|3}:=\ketbra{y_-}{y_-}.
   \label{eq:measDef3}
\end{align}
{Note that, due to the normalisation of probabilities, 
the measurements of some measurement operators are unnecessary. 
However, in practice, the orthogonal measurements $M_{2|j}$ are often measured to aid in the data normalization. We will include them here, with a view to our experiment.}

From these input states and measurements, one estimates the probabilities $p(a|x,w)=\Tr\Big(\mathcal{C}(\rho_w)\, M_{a|x}\Big)\,$, which can also be written as
\begin{align}
    p(a|x,w)=&\Tr\Big( C\, \left(\rho_w^T \otimes M_{a|x}\right)\Big),\\
    =&\Tr\left( C\, S_{a|xw} \right)
\end{align}
in the Choi formalism, where $S_{a|xw}:=\rho_w^T \otimes M_{a|x}$, where $S_{a|xw}$ is a setting operator for standard quantum process tomography.
Now, one way to perform complete tomography is by ensuring that the setting operators $S_{a|xw}$ span the space $\L(\H_\text{in}\otimes \H_\text{out})$. Also, thanks to a property usually referred to as ``local tomography''~\cite{wootters1990local} if the set of operators $\{\rho_w^T\}_w$ spans $\L(\H_\text{in})$ and the set $\{M_{a|x}\}_{a,x}$ spans $\L(\H_\text{out})$, then the set of setting operators $\left\{\rho_w^T\otimes M_{a|x}\right\}_{w,a,x}$ spans $\L(\H_\text{in}\otimes~\H_\text{out})$. In other words, full quantum channel tomography is always possible if one measures a set of characterised setting operators $\{S_{a|xw}\}_{a,x,w}$ that span the space $\L(\H_\text{in}\otimes~\H_\text{out})$. 

In principle, measuring a set of setting operators $\{S_{a|xw}\}_{a,x,w}$ which span $\L(\H_\text{in}\otimes \H_\text{out})$, is actually ``overkill''. More specifically, due to the normalisation condition $\Tr_\text{out}(C)=\id_\text{in}$, respected by quantum channels, there are linear operators in $\L(\H_\text{in}\otimes \H_\text{out})$ which cannot be written as linear combinations of quantum channels,\textit{e.g.,} $\ketbra{0}{0}_\text{in}\otimes \id_\text{out}$. One can then consider a set of operators $\{S_{a|xw}\}_{a,x,w}$ which spans the set of quantum channels, a subspace with dimension strictly smaller than the dimension of  $\L(\H_\text{in}\otimes \H_\text{out}) $. In particular, the linear space $\L(\H_\text{in}\otimes \H_\text{out}) $ has dimension of $d_\text{in}^2\,d_\text{out}^2$, and the linear span of quantum channels in $\L(\H_\text{in}\otimes \H_\text{out}) $ has dimension of $d_\text{in}^2(d_\text{out}^2-1)$. We emphasize, however, this does not represent a problem; in fact, in practice, more using an over-complete measurement set is known to minimise the experimental errors in standard quantum tomography \cite{deBurgh_choice_2008}.

\begin{figure*}
    \centering
    \includegraphics[width=\linewidth]{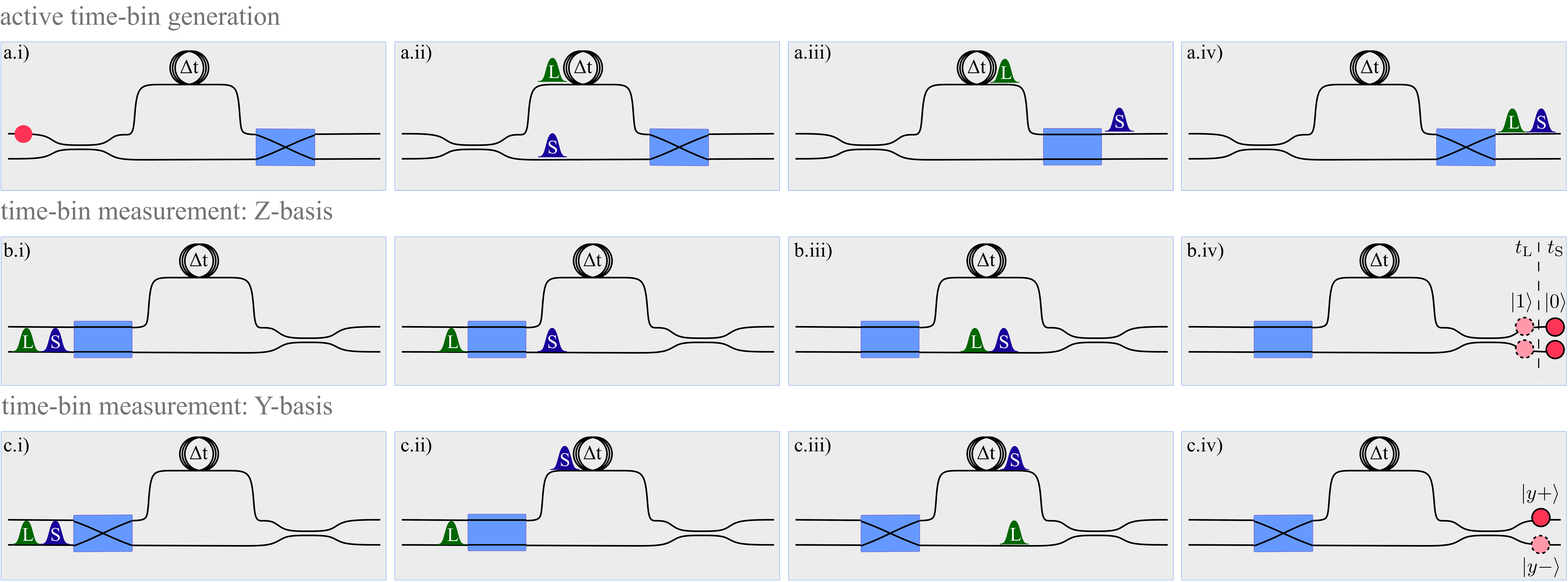}
    \caption{\textbf{Active generation and measurement of time-bin qubits.} a) In this row, an incident single photon is deterministically placed in a superposition of the ``short'' and ``long'' time bins, using an ultra-fast optical switch (UFOS).
    In order to achieve passive phase stability, the measurements, shown in rows b) and c), are actually implemented using the same device as in row a).
    However, for clarity, we have mirrored the device horizontally. b) This row indicates our measurements in the Z-basis.  Here, the UFOS remains in the ``bar state''. After traversing the device, the photon remains in a superposition of the two incident time bins (but spread over two paths).  In this situation, simply resolving the arrival time of the photon projects into the Z-basis.
    c) A schematic of our deterministic measurement in the Y-basis. Here the UFOS alternate between the ``cross'' and bar states so that the short (long) time bin now takes the long (short) path.
    In this manner, the two time bins interfere on the beamsplitter, so that finding the photon in the upper (lower) path corresponds to projecting the time bin onto $\ket{y+}$ ($\ket{y-}$).}
    \label{fig:time-bin generation}
\end{figure*}

Finally, we now consider tomography of process matrices $W\in~\L\left(\H_P \otimes \H_{A_\text{in}}\otimes\H_{A_\text{out}}\otimes\H_{B_\text{in}}\otimes\H_{B_\text{out}}\otimes\H_{F}\right)$, such as the quantum switch illustrated in Fig. \ref{fig::cartoon}d. As discussed before, one way to perform tomography is to measure setting operators $S_{abc|xyzw}$ which span the linear space $\L\left(\H_P \otimes \H_{A_\text{in}}\otimes\H_{A_\text{out}}\otimes\H_{B_\text{in}}\otimes\H_{B_\text{out}}\otimes\H_{F}\right)$. Also, thanks to local tomography, we may consider sets of states and measurements which span the local space individually. We then consider the set of states given by Eq.~\ref{eq:stateSet} and the set of measurements is given by Eq.~\ref{eq:measSet}.
For tomography of a {higher-order} process matrix, we then consider the set of measure-and-reprepare instruments {(to be used as inputs for Alice and Bob's channels)} given by all combinations of the two sets above, that is
\begin{align}
\label{eq:measrep1}
    \mathcal{R}:=\{R_{i|(j,k)}\}_{i=1,j=1,k=1}^{i=2,j=3,k=4},
\end{align} where
\begin{align}
\label{eq:measrep2}
    R_{i|(j,k)}:=M_{i|j} \otimes \ketbra{\psi_k}{\psi_k}^T.
\end{align}
{The interpretation of Eq.~\eqref{eq:measrep2} is the following, first, the measurement $j$ with POVM elements $\{M_{i|j}\}_i$ is performed, then, the state $\ket{\psi_k}$ is prepared. Notice that, in our measure-and-reprepare instruments, the prepared state $\ket{\psi_k}$ is independent of the measurement choice $j$ and the obtained outcome $i$. }
One can then perform full tomography with the setting operators
\begin{align}
\label{eq:settingOP}
    S_{abc|xyzw}&:={\ketbra{\psi_w}{\psi_w}^T}^{P_\text{t}}\otimes A_{a|x}^{A_I A_O} \otimes B_{b|y}^{B_IB_O} \otimes M_{c|z}^{F_\text{c}},
\end{align}
where $\ketbra{\psi_w}{\psi_w}$ are the $4$ different quantum states in $\mathcal{S}$, $A_{a|x}$ and $B_{b|y}$ are each the $2\times 3\times 4 =24$ instrument elements\footnote{Note that here we identify $x=(j,k)$, that is, $x$ can assume $12=3\times4$ different values.} of set $\mathcal{R}$ defined in Eq.~\eqref{eq:measrep1}, and $2\times3=6$ measurement elements for the future control space are the measurements $M_{c|z}$ of $\mathcal{M}$ (Eq.~\eqref{eq:measSet}).
In total, we then measure $4\times24\times24\times6=13824$ different settings.
We remark that in this tomography approach, we do not make use of any constraints on the process matrices. One could also to reduce the number of required settings by imposing the assumption that valid process matrices necessarily belong to a particular linear subspace, as discussed in subsection \ref{subsec:general_process}.

In an ideal theoretical scenario, if we obtain the probabilities $ p\left(abc|xyzw\right)=\Tr\left(W \,S_{abc|xyzw}\right)$ for a tomographically complete set of setting operators, standard linear inversion will uniquely identify the process $W$. However, due to finite statistics, we never obtain the exact probability $p\left(abc|xyzw\right)$, but an approximation from measured frequencies. Also, due to measurement precision and other possible sources of errors, we cannot expect to obtain an exact reconstruction of the process matrix. Indeed, performing direct linear inversion often results in unphysical quantum states or processes. 
Instead, we aim to estimate a physical process matrix that agrees best with the experimental data.

In order to estimate our experimental process matrix $W_\text{exp}$, we perform a fitting routine to find the process matrix that best describes our measured data.
We find that minimizing the least absolute residuals works quite well.
To do this, we numerically search for a process matrix $W_\text{exp}$ that minimizes the following expression:
\small
\begin{equation}
    r=\frac{1}{N_\text{Settings}}{\sum_{abcxyzw} \Big|p_{\text{exp}}(abc|xyzw)- \Tr(W_{\text{exp}}\, S_{abc|xyzw})}\Big|,
\label{eq::Minimization expression}
\end{equation} 
\normalsize
where $N_\text{Settings}$ is the number of setting operators, and the minimisation is further subject to the constraint that $W_\text{exp}$ is a valid process matrix. This minimisation can be performed by means of semidefinite programming (SDP), and may be implemented with the help of numerical libraries such as MOSEK. Our MATLAB code implementing this, is available at \cite{zenodoCODE}.
The first term under the root are the experimentally measured probabilities, while the second term corresponds to what is predicted by quantum theory for the characterised settings $S_{abc|xyzw}$. Since $W_{\text{exp}}$ is the only unknown quantity, the minimization of Eq.~\ref{eq::Minimization expression} delivers a process matrix that fits best to our experimental data, making no assumptions about the specific form of the process matrix.

\section{Experiment}

\subsection{Time-Bin Quantum SWITCH}
\label{sec:expSWITCH}
\begin{figure*}
    \centering
    \includegraphics[scale=0.16]{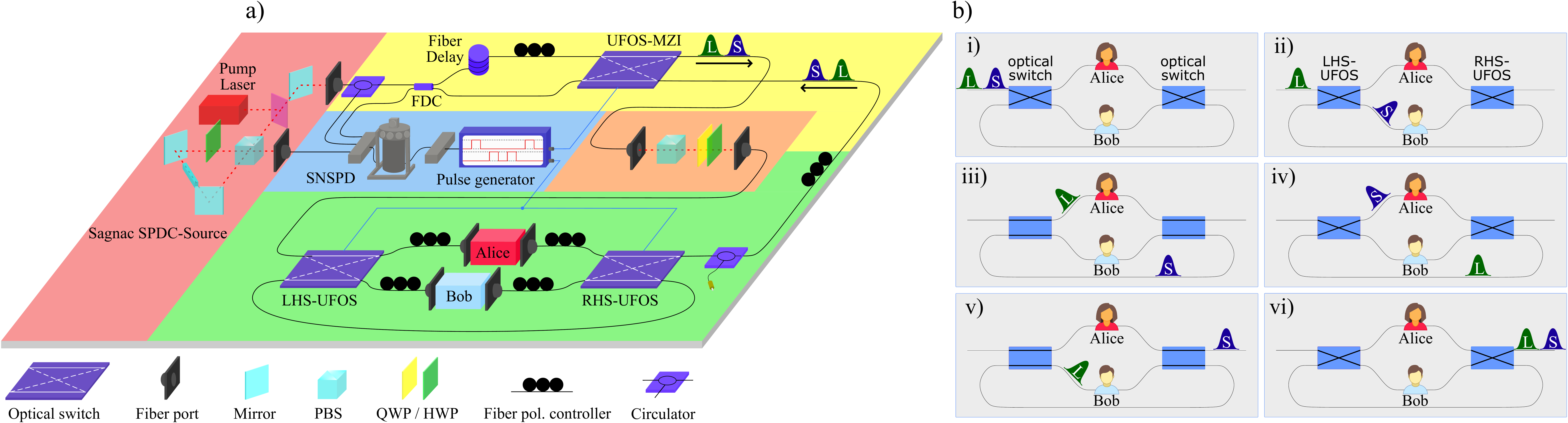}
    \caption{\textbf{Experimental setup.} \textbf{a)} The complete experiment. The individual sections are indicated with colors. The red section shows the Sagnac SPDC source that generates heralded single photons. In the yellow section, we illustrate the asymmetric MZI to generate and measure the time-bin control qubit. The orange section shows the target qubit preparation stage, which consists of a PBS and two waveplates. The green area hosts the fiber-based quantum SWITCH. The heralded and heralding photons are detected using SNSPDs, shown in the blue area. By triggering off of the detection event of a heralding photon we use a pulse generator to control the optical switches in the setup. The sub-panels \textbf{i) - vi)} in panel \textbf{b)} show the functionality of the quantum SWITCH. By controlling the state of the optical switches, we route the two time-bins in different orders through Alice and Bob's quantum channels. After the SWITCH operation, the target qubit has experienced the action of the quantum channels in a different order depending on the state of the time-bin qubit.}
    \label{fig:time-bin propagation}
\end{figure*}

To date, most previous implementations of the quantum SWITCH were based on bulk optics. Since photonic quantum SWITCHes are essentially interferometers, inevitable phase drifts limit the measurement time or require active stabilization \cite{Rubino2022experimental,Lorenzo}. 
Furthermore, since adding more parties means that the dimension of the control system must be increased, scaling up previous architectures requires more and more spatial modes to be transmitted through the same optic, making it difficult to create a SWITCH with more than two parties.
Here, we present a passively-stable, fiber-based architecture for the quantum SWITCH where the control system is encoded in a time degree of freedom of the photon.
Thus, in our architecture, although the dimension of the control system must still be increased at the same rate, this can be done using additional time bins, but only one spatial mode must traverse each optical element.
Furthermore, by using the same interferometer to prepare and measure the control system, all phase fluctuations cancel out, making our setup passively stable.
This is important for process tomography, as we must perform many measurements, and the experiment must remain stable during this time.

To create the time-bin qubit that we will use to control the order, we start by generating a photon pair, $\lambda=1550$ nm, using spontaneous parametric down conversion (SPDC). One photon of the pair is directly detected to herald the other photon, setting a time reference for the experiment. The second photon is sent to a 50/50 beamsplitter---a fiber directional coupler (FDC)---which splits the incoming mode into two fibers of different lengths. We then deterministically recombine these two fiber paths using an ultra-fast fiber optical switch (UFOS), see Fig. \ref{fig:time-bin generation}a (and also the yellow section of Fig. \ref{fig:time-bin propagation}a)~\cite{active_switching_Padova}. 
To do so, we generate an electronic pulse, triggered off of the timing reference generated by detecting the first photon.
This pulse is sent to the UFOS which change its state to first route the ``photon component'' from the short path, followed by the ``photon component'' in long path, into the upper output mode of the UFOS-MZI (Fig. \ref{fig:time-bin generation}a ii and iii). 
The result is that the second photon is left in an equal superposition of two time bins in a single fiber (Fig. \ref{fig:time-bin generation}a iv).
Note that because the short time bin is transmitted through the FDC, while the long time bin is reflected, one mode picks up a reflection phase, while the other does not.
Hence, in our experiment we prepare the control qubit in the state $\ket{y-}_\text{C}=(\ket{S}_\text{C}-i\ket{L}_\text{C})/\sqrt{2}$, where we have labelled the modes as the ``short'' (``long'') state $\ket{S}_{\text{C}}$ ($\ket{L}_{\text{C}})$, when it has taken the short (long) fiber path of the interferometer.
The spacing between these two time bins is $150$ ns, which is set by the response time of our UFOS. 

The UFOS we use to route the photon are BATi 2x2 Nanona fiber switches. In addition to creating the time bin qubit, they allow us to route the photon in a controlled way through the quantum SWITCH. Our UFOSs have a response time of 60 ns, with a maximal duty-cycle of 1 MHz, and a cross-channel isolation greater than 20 dB for any polarisation (see \cite{zanin2021fiber,zanin2022enhanced} for more details).

{
Having created the time bin control qubit, we need to apply the quantum SWITCH operation to the target system, which we encode in the polarization degree-of-freedom of the same photon.
To route the photon, we use two additional UFOSs and follow the protocol illustrated in Fig. \ref{fig:time-bin propagation}b.i-vi. 
In particular, we send a voltage pulse train consisting of three low levels and two high levels to the UFOS's. During each low level, the fiber switches are in a ``cross state'' (output modes are swapped with respect to the input), while during a high level the switch state is set to the ``bar state'' (input modes transmitted to output modes). As the time-bins approach the quantum SWITCH (Fig. \ref{fig:time-bin propagation}b.i) the UFOS's are initially in the cross-state, which routes the short time bin $\ket{S}_{\text{C}}$ through Bob's quantum channel (Fig. \ref{fig:time-bin propagation}b.ii). Then the UFOSs change to the bar state (Fig. \ref{fig:time-bin propagation}b.iii), which sends $\ket{L}_{\text{C}}$ through Alice's channel, while $\ket{S}_{\text{C}}$ travels over a fiber from the RHS-UFOS to the LHS-UFOS. Then the UFOSs see a low voltage level, and their state is set to cross (Fig. \ref{fig:time-bin propagation}b.iv). This sends $\ket{S}_{\text{C}}$ through Alice local laboratory, while $\ket{L}_{\text{C}}$ loops back to the LHS switch. In Fig. \ref{fig:time-bin propagation}b.v the UFOS's are in bar state and hence, $\ket{L}_{\text{C}}$ passes through Bob's channel. At this point $\ket{S}_{\text{C}}$ exits the quantum SWITCH. 
Finally, the fiber switches are set to the cross state (Fig. \ref{fig:time-bin propagation}b.vi) so that $\ket{L}_{\text{C}}$ leaves the quantum SWITCH. At this point, depending on the control state, the target system has experienced a different order of Alice and Bob's actions, which, as we will describe shortly, act on the polarization state of the photon. 
Note, that all the lengths of the fibers in the quantum SWITCH are set to ensure the correct routing of time-bins spaced by $150$ ns.}

{The time-bin quantum SWITCH from Fig. \ref{fig:time-bin propagation}b is placed in the full fiber-based setup (Fig. \ref{fig:time-bin propagation}a), in which the time-bins are prepared and measured.
The quantum SWITCH itself is shown in the green section of panel a). The type-II SPDC photon source\cite{SPDC_Sagnac} is shown in the red section\footnote{Our source is in a Sagnac configuration, although for this experiment we only pump the Sagnac loop in one direction so as not to generate polarization entanglement}. 
Here, a PBS reflects the heralding photon to a single photon detector (blue area), while the other photon is transmitted to the Mach-Zehnder-like time-bin generation interferometer explained above (yellow section).
Following this, we have a photon encoding a time-bin qubit in the state $(\ket{S}_{\text{C}}-i\ket{L}_{\text{C}})/\sqrt{2}$ in the ``upper'' output of the interferometer (clockwise direction). 
The counter-clockwise path of the loop (lower UFOS-MZI output mode) hosts an optical circulator with an empty port to filter out misguided photons, which can arise from the imperfect extinction ratio of our UFOSs.
Next, the target system is encoded in the photon's polarization. For this we use a polarizing beam splitter (PBS) and a set of a quarter- and a half-waveplate, shown in the orange section of Fig. \ref{fig:time-bin propagation}a. 
Then we apply the 2-SWITCH operation to the target system described above (green section of Fig.\ref{fig:time-bin propagation}a and Fig.\ref{fig:time-bin propagation}b), where we implement Alice and Bob's instruments using short free-space sections containing waveplates and polarizers.

After exiting the SWITCH, the photon follows the fiber loop in clockwise direction and approaches the MZI used for time-bin generation (yellow section); now, from the opposite direction in the lower path. At this point, we can decide to measure the control qubit in the computational ($Z$) or a superposition ($Y$) basis.
These measurements are illustrated in Fig. \ref{fig:time-bin generation}b and c, respectively.
For measurements in the computational basis, both time-bins are routed by the UFOS along the lower path of the MZI, after which the two time bins split up at the FDC, and are then sent to detectors in the blue region.
By measuring the arrival time, with respect to the herald detection, we can distinguish between the short and long time bins.
To measure in a superposition basis, we use UFOS-MZI to send the time bins through the opposite paths of the interferometer ($\ket{S}_C$ takes the long path and $\ket{L}_C$ takes the short path) so that they arrive at the FDC at the same time.
In this case, interference occurs at the FDC, and detecting a photon at exiting the upper (lower) port corresponds to projecting the control qubit in $\ket{y+}$ ($\ket{y-}$).
}

\begin{figure}
    \centering
    \includegraphics[width=\linewidth, trim=3cm 0cm 3cm 1cm, clip]{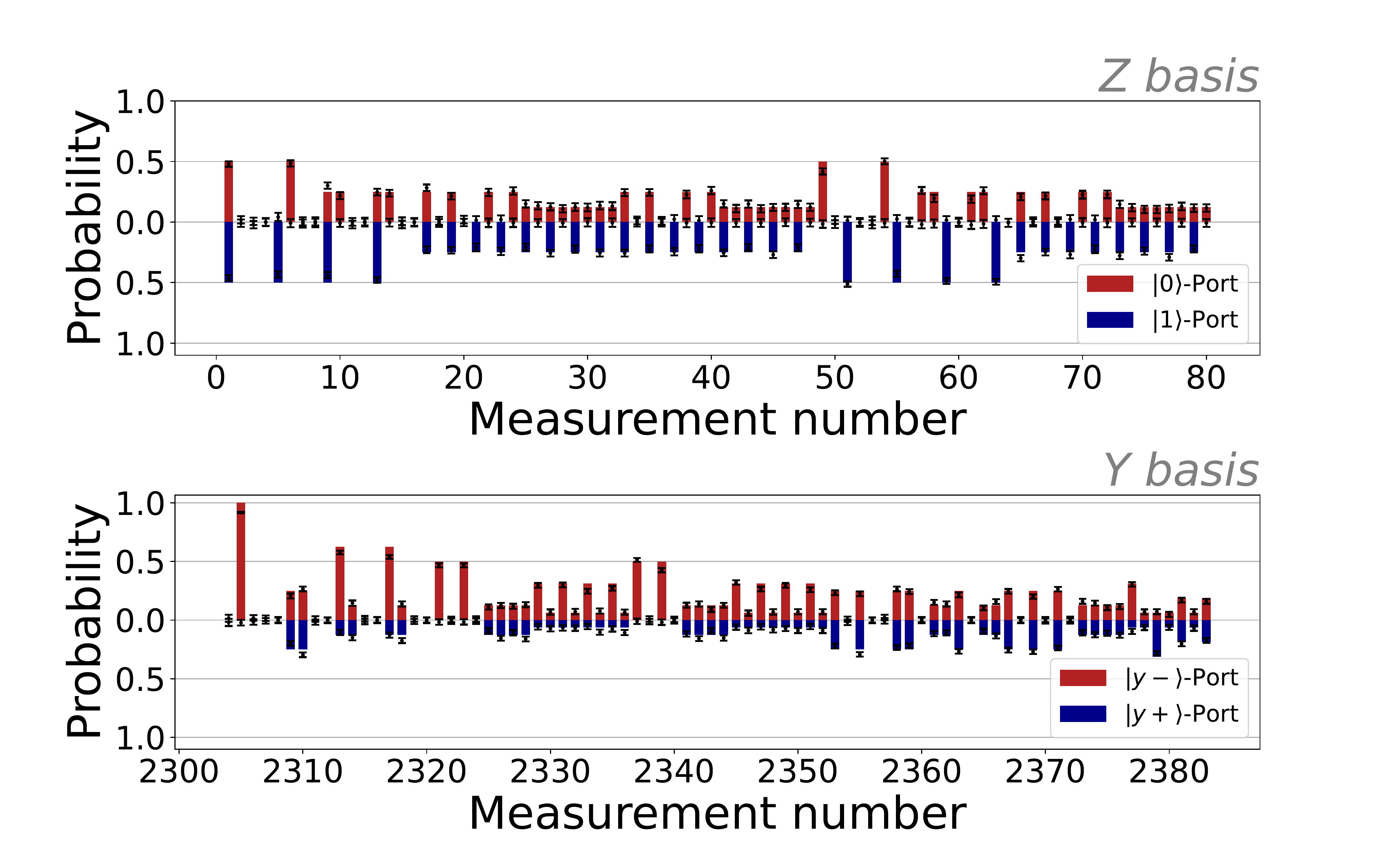}
    \caption{\textbf{Experimentally Estimated Probabilities}. {A small subset of the experimentally estimated probabilities. The bars are the theory for the ideal process matrix $W^{y-}$, and the points are the experimental estimates. The upper (lower) panel shows measurements of the control qubit in the $Z$ ($Y$) basis. 
    The red bars are for outcomes $\ket{0}$ and $\ket{y-}$, and the blue for outcomes $\ket{1}$ and $\ket{y+}$.}}
    \label{fig:Witness_probabilities}
\end{figure}

With this in place, we collect the measurement statistics from different measurement settings by detecting coincidence events between the heralding photon and the FDC output or the circulator output.
{For each experimental configuration, we record $\approx1600$ coincidence counts ($\approx21,000$ total single photon counts) over $10$ s at the FDC and circulator output. The photon source generates $\approx 1,480,000$ single photons ($\approx 116,000$ coincidence events) in $10$ s before the experiment. Thus, our entire quantum SWITCH experiment has an overall insertion loss of $\approx 18$ dB.}
All of our measurements are carried out with superconducting nanowire single photon detectors (SNSPD) from PhotonSpot Inc.
The result, for a representative set of measurements, is shown in Fig. \ref{fig:Witness_probabilities}.
Therein, the bars are the theory for an ideal quantum SWITCH with the control qubit in $\ket{y-}$, described by the process matrix $W^{y-}$, while the points are our experimentally measured data.
Already, one can observe good agreement between theory and experiment.
{Using unitary operations (rather than measure and reprepare instruments), we can also play the anti-commuting/commuting gate discrimination game, as in Ref. \cite{Lorenzo}.
We find a success probability of $0.974\pm0.018$, indicating a high-fidelity of our implementation (the full details of this measurement are presented in the Appendix \ref{sec::game}). }

\subsection{Experimental Process Matrix Tomography }
We will now present our experimental reconstruction of the process matrix of our time-bin quantum SWITCH. 
As discussed in Sec. \ref{sec:theory}, to probe the underlying process, Alice and Bob must each implement a complete set of instruments. 
In our experiment, the target system is encoded in the polarization state of the photon, so Alice and Bob must act on this degree of freedom.
Rather than the measurement-repreparation instruments defined in Eqs. \ref{eq:measrep1} and \ref{eq:measrep2}, we use a slightly modified form $\Tilde{R}:=\Tilde{R}_{i|(j,k)}$ presented in the Appendix Eqs. \ref{eq:measrepExp1} and \ref{eq:measrepExp2}.
In particular, Alice and Bob each have access to three different measurement bases $\Tilde{M}_{i|j}$ where $j \in {1,2,3}$ defines the measurement, and $i$ defines the outcome.
For each $j$,  $\Tilde{M}_j:=\Tilde{M}_{1|j}-\Tilde{M}_{2|j}$ is the observable associated with the POVM $\{\Tilde{M}_{1|j},\Tilde{M}_{2|j}\}$.
The specific operators we implement are defined in Eq. \ref{eq:measSetA} and \ref{eq:expMij}.
For example, $j=1$ corresponds to the $Z$ basis: $\Tilde{M}_1:=\Tilde{M}_{1|1}-\Tilde{M}_{2|1}=Z$.

Experimentally, we implement these measurements using a polarizer fixed to transmit horizontally polarized light $\ket{H}$. We set the measurement basis using a quarter waveplate and a half waveplate before the polarizer to the angles given in Eq. \ref{eq:expMij}.
To implement the second part of the instrument---the repreparations---we must prepare one of four different states.
We experimentally accomplish this using another quarter- and half-waveplate to rotate the photon's polarization if it is transmitted through the polarizer.
This allows us to prepare one of the four $\ket{\Tilde{\phi}_k}$ states listed in Eq. \ref{eq:expPhi}.
Thus, overall, both parties can implement the 24 different measurement-repreparation operators defined in Eqs. \ref{eq:measrepExp1} and \ref{eq:measrepExp2} (6 different measurement operators times 4 different repreparations)

In addition to Alice and Bob's channels, we must send in a complete set of target states, and perform measurements on the control qubit after the SWITCH.
To this end, we first prepare the target qubit in the four different input states $\ket{\Tilde{\psi}_w}$ given in Eq. \ref{eq:expPsi}.
We set these states using the quarter-half waveplate pair mounted in the target preparation stage, shown in the orange area of Fig. \ref{fig:time-bin propagation}a (the exact waveplate angles that we use are listed in Eq. \ref{eq:expPsi}).

Finally, at the output of the SWITCH we must measure the state of the control qubit.
This procedure is illustrated in Fig. \ref{fig:time-bin generation} panels b) and c).
As  discussed in Sec. \ref{sec:expSWITCH}, we use the same beamsplitter to measure and prepare the time-bin qubit, but from opposite directions. As a result, the phase of this measurement basis is fixed to the $Y$ basis.
In our notation in the Appendix Eqs. \ref{eq:expMeasSet} and \ref{eq:expContMeas}, this corresponds to a measurement $\Tilde{C}_{1|2}$ and $\Tilde{C}_{2|2}$.
Experimentally, a $\Tilde{C}_{1|2}$ versus $\Tilde{C}_{2|2}$ result depends on which port of the FDC the photon exits.
As described above, we can additionally measure in the $Z$ basis by fixing the UFOS-MZI to the bar state on the return trip such that the short and long time-bins do not interfere at the TDC and observing the arrival time of the time-bins. If we find the photon arrives earlier, this is associated with an $\Tilde{C}_{1|1}$ detection event, while if it arrives later corresponds to $\Tilde{C}_{2|1}$.

In order to be complete on the future control space, we would require an additional measurement of the control qubit in the $X$ basis, i.e. we need the measurements {$M_{1|2}$ and $M_{2|2}$} from Eq. \ref{eq:measDef2}. 
In our experiment, this could be achieved using a fast phase modulator to apply the appropriate phase between the short and long path only on the reverse direction.
However, we do not implement this here.
Instead, we impose an additional constraint on our tomographic reconstruction.
We require that $\Tr \left( W_\text{exp} {X}^{F}\right)= 0$; {where $X=\ketbra{+}{+}-\ketbra{-}{-}$}.
Given the passive phase stability of our experiment, this is a very good assumption. 
{We verify this assumption, by comparing reconstructions with and without this constraint.
In particular, we find that the fidelity between the process matrices reconstructed with and without this constrain is $0.999982$, well below our experimental error.}

Overall, this results in $24\times 24\times 4\times 4 = 9216$ setting operators of the form given in Eq. \ref{eq:settingOPEXP} (number of Alice's settings $\times$ number of Bob's settings $\times$ number of target states $\times$ number of control measurements).
However, for the control measurements, we have access to both ports of the FDC beamsplitter simultaneously (i.e there is a detector in each output port of the beamsplitter), giving rise to $4608$ different experimental configurations.
From these data we can calculate the probabilities for each given setting operator.
Experimentally, we measure count rates associated with each setting operator, which we must then normalize to convert into the required probabilities.
To do so, we make use of the normalization condition over the outcomes of all three measurements
\begin{align}
\sum_{abc} p(abc|xyzw) = 1.
\end{align}
Thus, we define a normalization constant for every value of $x$, $y$, $z$, and $w$
\begin{align}\label{eq:norm}
N_{xyzw}=\sum_{abc} C(abc|xyzw),
\end{align}
where $C(abc|xyzw)$ are the number of coincidences measured between the heralding detector and the detectors after the FDC, corresponding to the setting operator defined by $a$, $b$, $c$, $x$, $y$, $z$, and $w$.
Then our experimentally estimated probabilities are defined as 
\begin{align}
p_\text{exp}(abc|xyzw) = \frac{C(abc|xyzw)}{N_{xyzw}}.
\end{align}
A small subset of the resulting probabilities are plotted in Fig. \ref{fig:Witness_probabilities}.
Then, by minimizing Eq.~\ref{eq::Minimization expression} (with the setting operators $S_{abc|xyzw}$ replaced by the experimental setting operators $\Tilde{S}_{abc|xyzw}$ from Eq. \ref{eq:settingOPEXP}) we can reconstruct the process matrix $W_{\text{exp}}$.
{Our MATLAB code implementing this minimization is available at \cite{zenodoCODE}.}

\section{Results}
\subsection{Fidelity}
\label{sec:results:fidelity}
\begin{figure}[t]
    \centering
    \includegraphics[width=\linewidth, trim=15cm 2.5cm 15cm 2.5cm, clip]{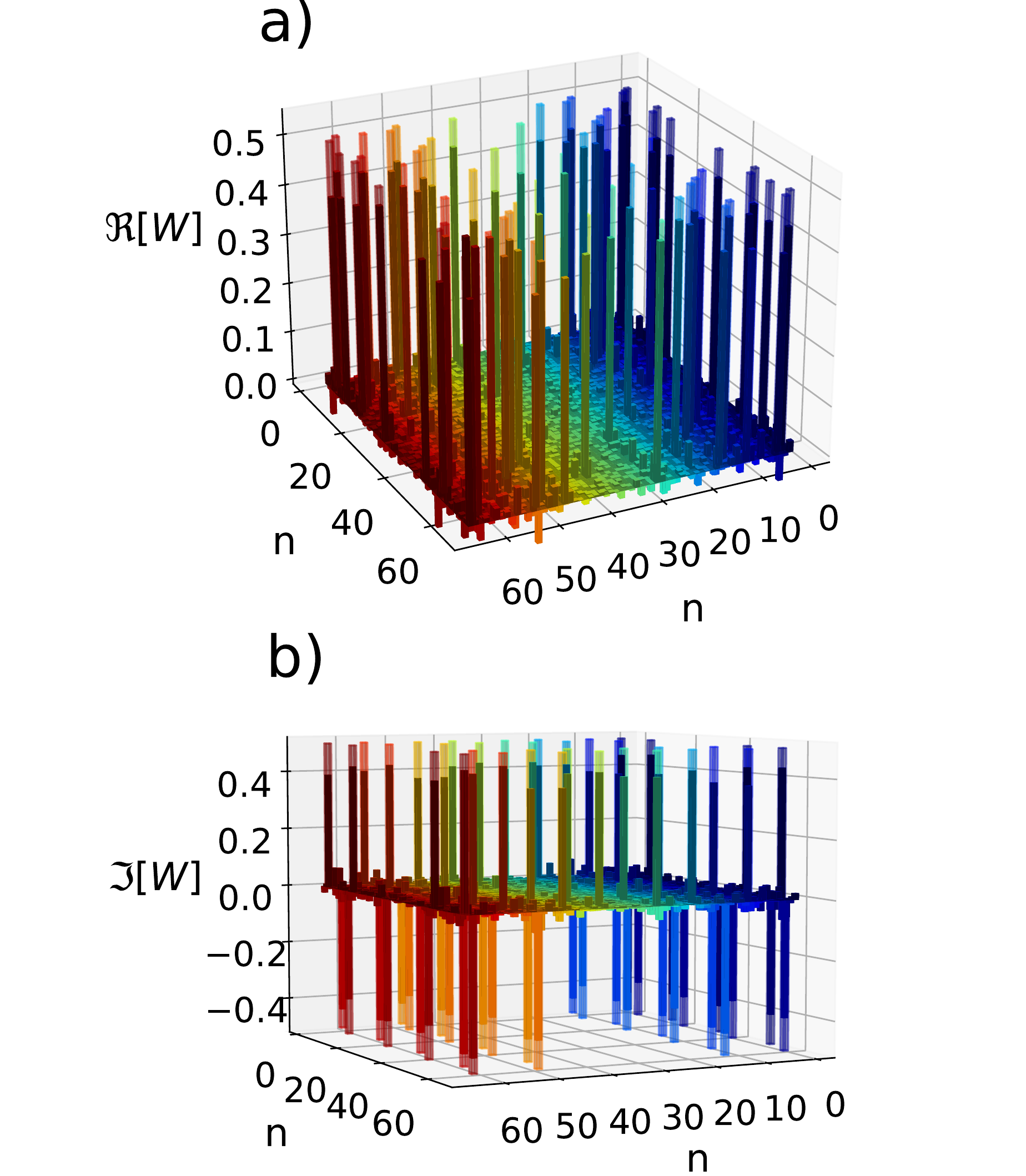}
    \caption{\textbf{Process tomography data.} This figure shows the experimentally recreated process matrix of the quantum SWITCH. \textbf{a)} represents the real part of $W_{\text{SWITCH}}$ and \textbf{b)} the imaginary part. 
    The color gradient along the x-axis does not have a physical meaning; rather, it is  color coded in order to identify the individual elements of this $64$x$64$ matrix better. Additionally, the ideal process matrix is represented via the layer of semi-transparent bars. Calculating the fidelity between these two process matrices results in $F(W_{\text{exp}}, W_{s}^{y-}) = 0.920\pm0.001$.}
    \label{fig:Data SWITCH process matrix}
\end{figure}

The experimentally obtained $64\times 64$ process matrix and the ideal matrix are plotted in Fig.~\ref{fig:Data SWITCH process matrix} as a 3D-bar chart, where panel a) shows the real part, and panel b) the imaginary part. The solid bars are the experimentally reconstructed process matrix, while the transparent bars are the theoretical process matrix $W^{y-}_s$.
The x and y axis numerically label the basis elements.
The relatively close agreement between the target process matrix and our experimental process matrix is already evident in this figure.

\begin{table*}
\begin{center}
\begin{tabular}{|c || c | c | c | c |}
 \hline & & & & \\[-2ex]
    & ${G}_\text{y-,res}^\text{GR}$ & ${G}_\text{exp,res}^\text{GR}$ & ${G}_\text{y-,all}^\text{GR}$ & ${G}_\text{exp,all}^\text{GR}$  \\ [0.5ex] 
 \hline\hline
 $W^{y-}_s$ & $-0.5834$ & $-0.5525$ &$-0.5834$ & $-0.5512$\\ 
 \hline
 $W_\text{exp}$&$-0.387\pm0.003$&$-0.431\pm0.003$ & $-0.370\pm0.003$ & $-0.431\pm0.002$\\
[1ex] 
 \hline
\end{tabular}
\caption{\textbf{Generalized Robustness Witness Analysis.} A summary of the different generalized robustness witnesses constructed.  The witnesses $G_{i,j}$ are labelled by two subscripts. The first indicates if the witness was designed for the ideal process matrix~$W^{y-}$ (subscript ''$y-$'') or the experimental process matrix $W_\text{exp}$ (subscript ''exp''). The second subscript indicates if the restricted measurement set (subscript ''res'') or the complete measurement set (subscript ''all'') was used for the witness. 
The first row shows the value of the witness for $W^{y-}$, and the second row shows the experimental values, which were evaluated as $\Tr[G_{i,j} W_\text{exp}]$. }
\label{tab:GR}
\end{center}
\end{table*}

\begin{table*}
\begin{center}
\begin{tabular}{|c || c | c | c | c |} 
 \hline & & & & \\[-2ex]
    & ${G}_\text{y-,res}^\text{WN}$ & ${G}_\text{exp,res}^\text{WN}$ & ${G}_\text{y-,all}^\text{WN}$ & ${G}_\text{exp,all}^\text{WN}$  \\ [0.5ex] 
 \hline\hline
 $W^{y-}_s$ & $-2.296$ & $-2.174$ &$-2.767$ & $-2.624$\\ 
 \hline
 $W_\text{exp}$&$-1.64\pm0.02$&$-1.76\pm0.01$ & $-1.96\pm0.02$ & $-2.112\pm0.02$\\
[1ex] 
 \hline
\end{tabular}
\caption{\textbf{White Noise Witness Analysis.} A summary of the different white noise witnesses constructed.  The witnesses $G_{i,j}$ and process matrix labels are labelled by two subscripts described in the caption of Tab. \ref{tab:GR}.}
\label{tab:WN}
\end{center}
\end{table*}

To further assess the agreement between our experiment and theory, we estimate the fidelity of the measured process matrix $W_{\text{exp}}$ to the ideal matrix $W_{s}^{y-}$. Since every valid process matrix normalized by its trace is a valid quantum state, we use the conventional expression for calculating the fidelity $F(\sigma, \rho) = \Tr\big(\sqrt{\sqrt{\sigma}\rho\sqrt{\sigma}}\big)$ with $\sigma$ and $\rho$ being different density matrices \cite{NielsenChuang}. This results in a fidelity of
\begin{equation}
    F(W_{\text{exp}}, W_{\text{SWITCH}}) = 0.920\pm0.001 \,,
\end{equation}
where the error arises is estimated using a Monte Carlo simulation of the entire reconstruction procedure accounting for finite measurement statistics and small waveplate errors of $1^\circ$.
Especially given the high-dimension of our process matrix, this fidelity indicates that our experiment is quite close to theory.

To quantify the agreement between our experimental data and $W_\text{exp}$ we compare the residuals of our fit~$r$ (defined in Eq. \ref{eq::Minimization expression}) to the average statistical error of our data $\eta_\text{stat}$.
The residuals $r$ can be interpreted as the disagreement between the  outcome predicted by $W_\text{exp}$ and the measured experimental outcome, averaged over all measurement settings.
For our fit $r=0.0089$, indicating an excellent match to our experimental data.
We estimate our statistical errors as follows.
First, we treat the probability to obtain an outcome $abc$ as a binomial variable: either we detect a photon or we do not.  Then we estimate the variance of that setting as $N_{xyzw}~p(abc|xyzw)\times(1-p(abc|xyzw))$, where $N_{xyzw}$ is the number of photons detected in all outcomes associated with $xyzw$ (defined in Eq. \ref{eq:norm}).  Finally, we compute the average error per setting as 
\begin{align}
    \eta_\text{stat}=\frac{1}{N_\text{settings}}\sum_{abc|xyzw}\frac{p(abc|xyzw)(1-p(abc|xyzw)\pink{)}}{\sqrt{C}}.
\end{align}
This is simply the standard error of each setting operator averaged over all settings.
Evaluating this for our data, we find $\eta_\text{stat}=0.0056$.
Given that $\eta_\text{stat}\approx r$, we conclude that our process matrix fits our data well.

\subsection{Causal Non-Separability}
A bipartite process matrix without a common past is causally non-separable when it cannot be written as a classical mixture of causally ordered processes~\cite{oreshkov2012quantum,Arajo2015}.
When considering bipartite processes with a common past, such as the quantum switch considered in this work, there are different non-equivalent definitions of indefinite causality. 
In Ref.~\cite{Arajo2015,rubino2017}, a bipartite process matrix $W\in~\L\left(\H_P \otimes \H_{A_\text{in}}\otimes\H_{A_\text{out}}\otimes\H_{B_\text{in}}\otimes\H_{B_\text{out}}\otimes\H_{F}\right)$  with common past and common future is said to be causally separable if it can be written as a convex sum of causally-ordered process matrices. That is, if we can write
\begin{align} \label{eq:def1}
    W = p W^{A>B} + (1-p) W^{B>A},
\end{align}
where $p\in[0,1]$, and $W^{A>B}$ and $W^{B>A}$ are causally-ordered process (objects also referred to as quantum combs~\cite{chiribella09Networks,Chiribella2013}, see Appendix~\ref{app:def_indefinite}). Alternatively, Ref.~\cite{giarmatzi15_causallySEP} proposes the concept of extensible causally-separable processes. This leads to a definition which differs from the one in Eq.~\eqref{eq:def1}, but is equivalent to the definition of causal-separability presented in Ref.~\cite{wechs2019_definition}, which considers not only convex mixtures of causally-ordered processes, but also incoherent (hence, classical) control of causal order. 
The analysis and the numbers presented in this section and in the main part of this paper were obtained via the definition of ~\cite{rubino2017}, which is the one presented in Eq.~\eqref{eq:def1}. However, we stress that the results of our work are not qualitatively affected by the different before-mentioned definitions, in the sense that, in all cases, the process we obtain after tomography is not causally separable, and it is robust against different kinds of noise. In Appendix~\ref{app:def_indefinite}, we present a more detailed discussion of such definitions and how they make small quantitative changes in the numbers presented here.

One method to quantify the degree to which our quantum process is causally non-separable is by using a causal witness.
A causal witness is a Hermitian non-negative operator ${G}$ such that $\Tr \left( {G} W_\text{sep}\right)\geq 0$ for all causally-separable processes. However, for all causally--non-separable processes (such as the quantum SWITCH) one can always find a witness ${G}$ such that $\Tr \left( {G} W^{y-}_s \right) <~0$.
Without additional constraints, the quantity $\Tr \left( {G} W \right)$ does not have a physical meaning, and may be artificially inflated by multiplying the witness $G$ by some constant. However, by setting additional normalisation constraints on the witness $G$, one may identify the quantity  $\Tr \left( {G} W \right) $ with how much noise the process $W$ can tolerate until it becomes causality separable. 
More concretely, let $\id_W:=\frac{\id}{d_P d_{A_O}d_{B_O}}$ be the ``white noise process'', which simply consists of discarding everything and outputting white noise, and let $W_\text{sep}$ be an arbitrary causally-separable process matrix.  Ref.~\cite{Arajo2015} shows that the problem
\begin{align}
    &\min \Tr(G W),\\
    &\text{s.t. } \Tr \left( {G} W_\text{sep}\right)\geq 0 , \quad \forall W_\text{sep} \\ 
    & \Tr(G)\leq \Tr(\id_W)
\end{align}
is equivalent to its dual formulation
\begin{align}
    &\min -r,\\
    &\text{s.t. } \frac{W+r\id_W}{1+r} \text{ is causally separable }.
\end{align}
Hence, under the normalisation constraint $\Tr(G)\leq~\Tr(\id_W)$, we ensure the identity $\Tr(G W)=-r$, which allows us to interpret $\Tr(G W)$ as how robust $W$ is against white noise.

Alternatively, one may also consider the normalisation $\Tr(G \Omega)\leq 1$, where $\Omega$ is an arbitrary process matrix. In this case, the equivalent problem reads as
\begin{align}
    &\min -r,\\
    &\text{s.t. } \frac{W+r\Omega}{1+r} \text{ is causally separable },
\end{align}
where $\Omega$ is an arbitrary process matrix. In this case, the value $\Tr(GW)=-r$ is typically called the ``generalised robustness''; it may be viewed as the amount of noise one needs to add to $W$ to make it causally separable in the worst case scenario.

In order for the witness $G$ to fit the setting operators implemented in our experiment, we impose an additional structure on the witness $G$ which is given by
\begin{equation}
\label{eq:wit}
    {G}= \sum_{abcxwyz}
    \alpha_{a,b,c,x,y,z,w} S_{abc|xyzw},
\end{equation}
where $\alpha_{a,b,c,x,y,z,w}$ are arbitrary real numbers and  $S_{abc|xyzw}$ are the setting operators of our experiment (see Sec.~\ref{sec:meas}). Additionally, for fixed setting operators,  finding the maximal violation of a witness $G$ with the normalization constraints related to white and generalised noise is a Semidefinite Program (SDP) \cite{Arajo2015}, and can be efficiently solved numerically~\cite{boyd}.
{Our code doing so is also available at \cite{zenodoCODE}.}

With these tools, we can construct a variety of witnesses.
First, we can construct witnesses using the complete measurement set (Eq. \ref{eq:settingOP}) or our restricted measurement set (Eq. \ref{eq:settingOPEXP}).
We can further design witnesses for two different process matrices $W^{y-}_s$ or $W_\text{exp}$.
This results in four witnesses:
${G}_\text{y-,all}$, ${G}_\text{y-, res}$, ${G}_\text{exp,all}$, ${G}_\text{exp, res}$.
Where the subscript $y-$ (exp) indicates that the witness was optimized for the ideal (experimental) process matrix, and the subscript all (res) indicates that the witness was computed using the complete (restricted) measurement set.
We can then further construct witness for either the generalized or white noise robustness.

The results of the generalized robustness witnesses are summarized in Tab. \ref{tab:GR}.
The first row of Tab. \ref{tab:GR} shows the value of the four witnesses evaluated using $W^{y-}_s$, and the second row shows the experimental values, estimated using $W_\text{exp}$.
In this case, we see that $\Tr(W^{y-}G_\text{y-,res}^\text{GR})\approx~\Tr(W^{y-}G_\text{y-,all}^\text{GR})$ and $\Tr(W^\text{exp}G_\text{exp,res}^\text{GR})\approx~\Tr(W^\text{exp}G_\text{exp,all}^\text{GR})$.
In other words, the generalized robustness evaluated either with the complete or restricted setting operators is equal within experimental error.
Evidently, the additional measurement on the future control system does not affect the generalized robustness.
More interesting for the generalized robustness witnesses, is the performance of the witnesses optimized for our experimental process matrix $W_\text{exp}$.
Examining the performance of our experimentally estimated witnesses (the bottom row of Tab. \ref{tab:GR}), we see that the witnesses designed specifically for our experimental process matrix increase the generalized robustness.
In particular,
$\Tr(W_\text{exp}G_\text{exp,all}^\text{GR})>~\Tr(W_\text{exp}G_\text{y-,all}^\text{GR})$ 
and $\Tr(W_\text{exp}G_\text{exp,res}^\text{GR})>~\Tr(W_\text{exp}G_\text{y-,res}^\text{GR})$.
This would not be readily possible without performing process matrix tomography.

In Tab. \ref{tab:WN} we summarize the results of our white noise witness analysis.
In this case, we see a significant difference between the witnesses constructed with the restricted and complete measurement sets, with the complete measurements sets revealing a higher white noise robustness in all cases.
Furthermore, in the second row, we can see that each step progressively improves the experimental white noise robustness.
The first entry $\Tr(G_\text{y-,res}^\text{WN} W_\text{exp})=-1.65\pm0.02$ shows the value that one would obtain for our setup without performing process matrix tomography. i.e. the witness was designed for the ideal process and uses only the experimentally implementable measurement settings.
In the next column, $\Tr(G_\text{exp,res}^\text{WN} W_\text{exp})=-1.76\pm0.01$ is improved by tailoring the witness for our experiment; however, still using only experimentally implementable measurements.
In the next two columns, we improve both of these values further by computing the witness assuming the complete measurement set.
The final entry, $\Tr(G_\text{exp,all}^\text{WN} W_\text{exp})=-2.11\pm0.02$ is significantly higher than the first entry, clearly showing the power of full process matrix tomography. 
Process matrix tomography allows us to compute properties of the experimental process, without having direct experimental access to them and we can precisely tailor our analysis to our experimental conditions.
In the Appendix tables \ref{tab:GR2} and \ref{tab:WN2}, we show the same analysis, for the alternative definition of causal non-separability.  The trends observed therein are the same, although the absolute values of the robustnesses are smaller.

\subsection{Worst-Case Process Tomography}

\begin{figure}
    \centering
    \includegraphics[width=\linewidth, trim=0cm 3cm 0cm 1cm, clip]{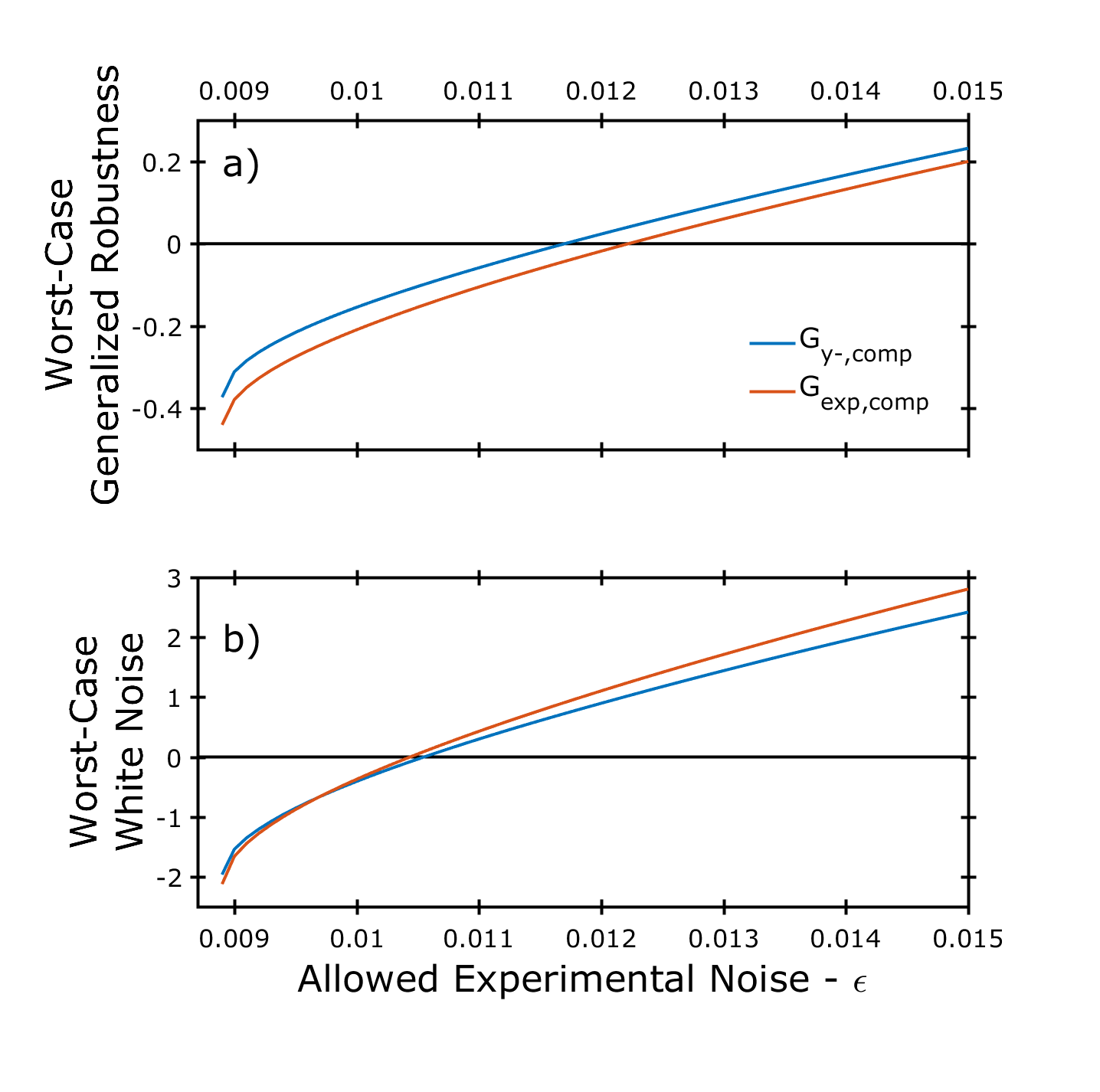}
    \caption{\textbf{Worst-Case Process Tomography}. Plots of the worst-case generalized robustness (a) and white noise robustness (b) causal witnesses versus the allowed deviation from the experimental data.  For these plots, the tomography routine attempted to find a process matrix which maximized the witness (i.e. it searched for the ``most causally separable'' process matrix), while still agreeing with our experimental data with an average error of $\epsilon$.  For all witnesses considered, the process which minimizes $\epsilon$ is also the most causally non-separable. For $\epsilon<0.089$ no valid process matrix is found.\\ }
    \label{fig:worstCaseRestults}
\end{figure}

For the process tomography results presented in Sec.~\ref{sec:results:fidelity}, we found the process matrix that fit to our data best, by minimizing Eq. \ref{eq::Minimization expression}. As discussed there, this resulted in a process matrix that describes our data very well.
However, one could ask ``Are there other causally-separable process matrices that describe the data almost as well?''
To answer this question, we perform a ``worst-case'' version of our process matrix tomography.
To do so, rather than minimizing expression Eq. \ref{eq::Minimization expression}, we find the process matrix that maximizes the generalized or white noise robustness, 
$\Tr(W_\text{worst} {G}_\text{j,all}^\text{GR})$ or $\Tr(W_\text{worst} {G}_\text{j,all}^\text{WN})$, respectively.
We do this using the witnesses designed for the ideal process matrix and the originally reconstructed experimental process matrix, but always with the complete measurement sets.
This maximisation is subject to the constraints that $W_\text{worst}$ is a physical process matrix and that the predictions of $W_\text{worst}$ match the experiment within some error $\epsilon$:
\begin{align}
    \sum_{abcxyzw}\abs{\Tr(S_{abc|xyzw} W)-p_\text{exp}(abc|xyzw)}\leq \epsilon.
\end{align}
Here $\epsilon$ is closely related to the residuals defined in Eq.~\ref{eq::Minimization expression}.
In particular, if $\epsilon<r$ the maximization will fail, as there is no physical process matrix compatible with this constraint.
Thus, we perform worst case process tomography for the four witnesses discussed above starting from $\epsilon=r_{\text{exp}}=0.0089$ and increasing to $\epsilon=0.015$.
The results of this analysis are plotted in Fig. \ref{fig:worstCaseRestults}a and b.
We find that the generalized robustness witnesses are more tolerant $\epsilon$, finding that our data is only consistent with causally separable process matrices for $\epsilon\lesssim 0.012$, while the white noise witnesses require $\epsilon\lesssim 0.0105$.
Although this analysis suggests that the causal non-separability is rather fragile, we stress the worst-case nature of this treatment: if a single causally separable process matrix is compatible to our data within $\epsilon$ it will be returned, even if a causally non-separable data fits our data better.
In any case, we see that for a range of experimentally relevant errors our data are only compatible with a causally non-separable process matrix.

\section{Discussion}
In this work, we have presented a protocol to perform process tomography on a higher-order quantum operation, the quantum SWITCH.
We discussed how to construct a complete set of measurements.
The requirements for this go beyond standard quantum process tomography, wherein one must ``only'' send a complete set of input states through the process, and perform a complete set of measurements after the process.
In particular, because HOQOs take quantum channels as inputs, we must also implement a complete set of quantum channels for each input channel.
This can be achieved using measure-and-reprepare instruments.
Since this procedure scales even worse than standard process tomography, we implement it using a new phase-stable architecture of the quantum SWITCH, allowing for long integration times.

Our photonic quantum SWITCH uses a time-bin qubit as the control system.
By recombining time-bin qubits using a passively-stable interferometer, we were able to keep our experiment stable indefinitely.
We believe this technique will be beneficial for various time-bin quantum information experiments and may even be scaleable to high dimensional time-bin qudits.
This would enable the construction of a multi-party quantum SWITCH.
The results of performing quantum process matrix tomography on our experiment show that we have indeed implemented a high-fidelity quantum SWITCH, with a fidelity of $F=0.920\pm0.001$. 
We then used our results to verify the causal non-separability of our experiment, designing causal witnesses specifically for our experimental process matrix.
Finally, we implemented a worst-case process tomography, searching for a causally-separable process that could also describe our measurement. To find such a process, we had to allow for a $\approx 1.5$ times larger disagreement between our measurements and our causally separable model.

Although our protocol was presented for the quantum SWITCH, it could be adapted to general HOQOs in a straight-forward manner.
In our present work, we performed an over complete set of measurements, but it should be possible to implement a reduced set of measurements by taking into account the constraints on the space of physical process matrices.
We also point out that many complexity-reducing techniques from standard state and process tomography, including compressed sensing \cite{flammia2012quantum}, shadow tomography \cite{aaronson2018shadow}, adaptive tomography \cite{mahler2013adaptive}, \textit{etc}., should apply to our protocol equally well.
But we leave these as topics for future work.

\section*{Data Availability}
{All the data that are necessary to replicate, verify, falsify and/or reuse this research is available online at \cite{zenodoCODE}.}

\section*{Acknowledgements}
This project was funded in whole, or in part, from the European Union’s Horizon 2020 research and innovation programme under grant agreement No 820474 (UNIQORN) and No 899368) (EPIQUS), the Marie Skłodowska-Curie grant agreement No 956071 (AppQInfo), and the QuantERA II Programme under Grant Agreement No I 6002 - N (PhoMemtor);
the Austrian Science Fund (FWF) through [F7113] (BeyondC), and [FG5] (Research Group 5);
the AFOSR via FA8655-20-1-7030 (PhoQuGraph), and FA9550-21- 1-0355 (QTRUST);
the Austrian Federal Ministry for Digital and Economic Affairs, the National Foundation for Research, Technology and Development and the Christian Doppler Research Association. For the purpose of open access, the author has applied a CC BY public copyright licence to any Author Accepted Manuscript version arising from this submission.

\bibliography{main}

\appendix
\section{Experimental techniques}\label{app:exp}

In this section, the used SPDC photon pair source is described and the technique implemented to compensate for unwanted polarization transformations induced by the birefringence of the optical fibers is explained. Moreover, the implemented experimental instruments and measurements are shown in detail.

\subsection{Photon Source}
\label{app::photon source}
To generate photon pairs at a wavelength of $\lambda = 1550$ nm we use a 775 nm, CW laser beam to pump a 30mm long ppKTP crystal. A dichroic mirror reflects 775 nm light and transmits photons at 1550 nm. The signal and idler photon are then separated on a PBS and coupled over optical fibers into the setup.

\subsection{Polarization Compensation}
In order to ensure that the quantum SWITCH performs the desired transformations on the photon's polarization, it is important to correct for the birefringent behaviour of the fibers that connect Alice and Bob's laboratories; i.e. to ensure that the fibers do not change the polarization state of the photon. Hence, each fiber link has to perform an identity operation. 
To this end, each fiber is equipped with a 3-loop fiber polarization controller, which allows us to implement any unitary polarization transformation in the fiber. In order to implement a true identity operation, we must check that the correct transformation is applied in two different bases. 
We use the computational and diagonal bases. 
A convenient way to do this, is to send classical light at the same wavelength as the single photons through the fibers and detect the polarization with a polarimeter at the fiber output. To ensure the identity transformation, we switch the light's polarization state at the target preparation stage(see Fig.~\ref{fig:time-bin propagation}) between horizontal $\ket{H}$ and diagonal $\ket{D}$, while adjusting the polarization controller until it converges to the correct setting in both bases.
To correct the polarization for the second trip through the channels, we place a polarizer in one of the channels (without additional waveplates). This decouples the compensation from the previous fiber. Then, we follow the same procedure and alternate the polarizer to transmit $\ket{H}$ and then $\ket{V}$. 
In order for this procedure to work properly, it is essential that wavelengths of the classical light and the single photons is matched.

\subsection{Experimental Instruments and Measurements}
Here we explain in detail the measurements and instruments we implement in the lab, and how they relate the ideal settings discussed in the main text.

\subsubsection{Future Control Measurement}
Our control qubit is the time-bin qubit. Its past state is fixed to $\ket{y-}$.
Ideally, we would measure the future control in three different bases described by Eqs.~\ref{eq:measDef}~-~\ref{eq:measDef3}.
However, due to experimental limitations we only measure the future control in two bases 
\begin{align}
\label{eq:expMeasSet}
    \mathcal{\Tilde{C}}:=\{\Tilde{C}_{c|z}\}_{c=1,z=1}^{c=2,z=2},
\end{align}
where
\begin{align}
\label{eq:expContMeas}
\begin{tabular}{|c | c || c|} 
 \hline & & \\ 
    c & z & $\Tilde{C}_{i|j}$  \\ [1ex] 
 \hline\hline
 $1$ & $1$ & $\ketbra{1}{1}$ \\ [1ex] 
 \hline
 $2$& $1$ &  $\ketbra{0}{0}$ \\ [1ex] 
 \hline
 $1$& $2$ &  $\ketbra{y-}{y-}$ \\ [1ex] 
 \hline
 $2$& $2$ & $\ketbra{y+}{y+}$ \\ [1ex] 
 \hline
\end{tabular}
\end{align}
Experimentally, the measurement outcomes $c=1$ and $c=2$ correspond to finding the photon exiting different ports of the beamsplitter (labeled FDC in Fig. \ref{fig:time-bin propagation} ).  Note that the order of the output indices has swapped compared to Eqs. \ref{eq:measDef}-\ref{eq:measDef3} in order to be consistent with our experimental convention.

\subsubsection{Past Target States}
Our target system is encoded as in the polarization degree of freedom of the photon.
We thus prepare its state by sending the photon to a polarizer set to transmit horizontal polarization, which we define to be the logical $\ket{0}$ state. We then set its state using a quarter waveplate, followed by a half waveplate.
We can thus prepare the set of states given by 
\begin{align}
\label{eq:stateSetEXP}
    \mathcal{\Tilde{S}}:=\{\ketbra{\Tilde{\psi_w}}{\Tilde{\psi}_w}\}_{w=1}^4,
\end{align} 
where
\begin{align}
\label{eq:expPsi}
\begin{tabular}{|c | c | c || c|} 
 \hline & & & \\ 
    QWP & HWP & w & $\ket{\tilde{\psi}_w}$  \\ [1ex] 
 \hline\hline
 $0^\circ$ & $0^\circ$ & $1$ & $\ket{0}$ \\ [1ex] 
 \hline
 $0^\circ$& $-45^\circ$ & $2$ & $\ket{1}$ \\ [1ex] 
 \hline
 $0^\circ$& $-22.5^\circ$ & $3$ & $\ket{-}$ \\ [1ex] 
 \hline
 $-45^\circ$& $0^\circ$ & $4$ & $\ket{y+}$ \\ [1ex] 
 \hline
\end{tabular}
\end{align}


\subsubsection{Alice and Bob's Instruments}
As our target system is encoded in a polarization state, Alice and Bob must implement measure and re-prepare channels on this degree of freedom.
To do the measurement, they use a fixed polarizer to project onto horizontal polarization.  Using a quarter and a half waveplate before the polarizer, they can then set the measurement basis.
Since this only provides one outcome (either the photon is transmitted or not, but they cannot  detect the cases when a photon is absorbed by the polarizer) they must also explicitly set the waveplates to perform the orthogonal measurement in order to normalize the data to compute a probability.
These measurements defined by the set
\begin{align}
\label{eq:measSetA}
    \mathcal{\Tilde{M}}:=\{\Tilde{M}_{i|j}\}_{i=1,j=1}^{i=2,j=3},
\end{align}
where the waveplates angles and resulting measurement operators are given by
\begin{align}
\label{eq:expMij}
\begin{tabular}{|c | c | c | c || c|} 
 \hline & & & & \\ 
    QWP & HWP & $i$ & $j$ & $\Tilde{M}_{i|j}$  \\ [1ex] 
 \hline\hline
$0^\circ$ & $0^\circ$ & $1$ & $1$ & $\ketbra{0}{0}$\\ [1ex] 
 \hline
 $0^\circ$& $45^\circ$ & $2$ & $1$ & $\ketbra{1}{1}$\\ [1ex] 
 \hline
 $45^\circ$& $22.5^\circ$ & $1$ & $2$ & $\ketbra{+}{+}$\\ [1ex] 
 \hline
 $45^\circ$& $67.5^\circ$ & $2$ & $2$ & $\ketbra{-}{-}$\\ [1ex] 
 \hline
 $45^\circ$& $0^\circ$ & $1$ & $3$ & $\ketbra{y-}{y-}$\\ [1ex] 
 \hline
 $45^\circ$& $45^\circ$ & $2$ & $3$ & $\ketbra{y+}{y+}$\\ [1ex] 
 \hline
\end{tabular}
\end{align}

Following their measurements, Alice and Bob reprepare the target state in one of 4 different options 
\begin{align}
\label{eq:stateRepEXP}
    \mathcal{\Tilde{P}}:=\{\ketbra{\Tilde{\phi_k}}{\Tilde{\phi}_k}\}_{k=1}^4.
\end{align}
This is again accomplished with a quarter and half wavplate.  Since the their polarizers transmit horizontal polarization, the horizontally polarized post-measurement photon is then rotated to one of the following states:
\begin{align}
\label{eq:expPhi}
\begin{tabular}{|c | c | c || c|} 
 \hline & & & \\ 
    QWP & HWP & k & $\ket{\tilde{\phi}_k}$  \\ [1ex] 
 \hline\hline
 $0^\circ$ & $0^\circ$ & $1$ & $\ket{0}$ \\ [1ex] 
 \hline
 $0^\circ$& $45^\circ$ & $2$ & $\ket{1}$ \\ [1ex] 
 \hline
 $0^\circ$& $22.5^\circ$ & $3$ & $\ket{+}$ \\ [1ex] 
 \hline
 $45^\circ$& $0^\circ$ & $4$ & $\ket{y-}$ \\ [1ex] 
 \hline
\end{tabular}
\end{align}

The net action of their instruments is then given by all combinations of the sets $\mathcal{\Tilde{M}}$ and $\mathcal{\Tilde{P}}$, that is
\begin{align}
\label{eq:measrepExp1}
    \mathcal{\Tilde{R}}:=\{\Tilde{R}_{i|(j,k)}\}_{i=1,j=1,k=1}^{i=2,j=3,k=4},
\end{align} where
\begin{align}
\label{eq:measrepExp2}
    \Tilde{R}_{i|(j,k)}:=\Tilde{M}_{i|j} \otimes \ketbra{\Tilde{\phi}_k}{\Tilde{\phi}_k}^T.
\end{align}

We then have the following experimental setting operator
\begin{align}
\label{eq:settingOPEXP}
    \Tilde{S}_{abc|xyzw}&:={\ketbra{\Tilde{\psi}_w}{\Tilde{\psi}_w}^T}^{P_\text{t}}\otimes \Tilde{A}_{a|x}^{A_I A_O} \otimes \Tilde{B}_{b|y}^{B_IB_O} \otimes \Tilde{C}_{c|z}^{F_\text{c}},
\end{align}

\section{Additional Measurements}

\begin{figure}[t]
    \centering\includegraphics[width=\columnwidth, trim=2cm 0cm 2cm 0cm,clip]{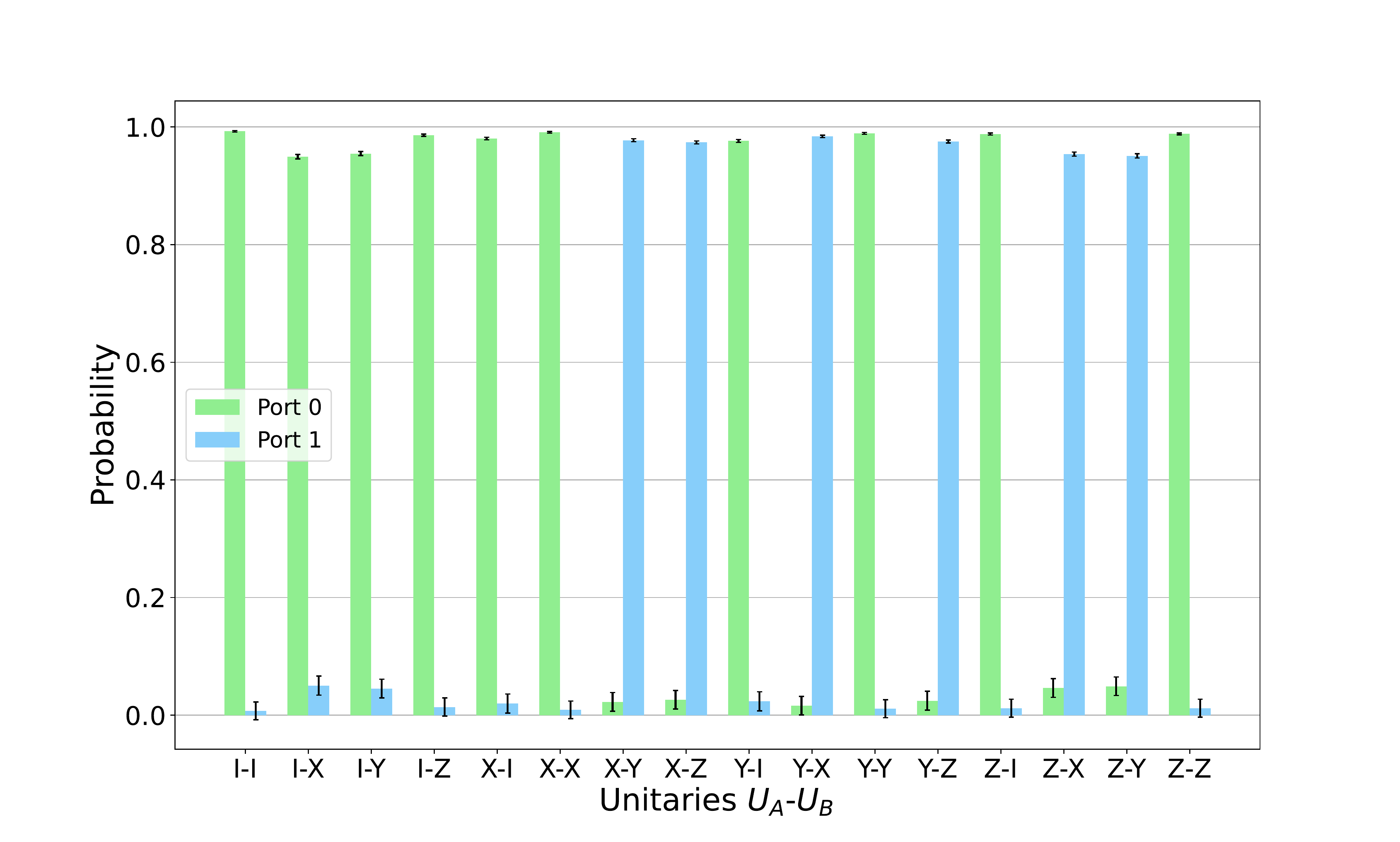}
    \caption{\textbf{Commutation Game.} This figure shows the experimentally estimated probabilities for the commutation / anti-commutation game. The green bars represent the probability, that the photon leaves port 0, leading the conclusion that the unitaries commute. The blue bars show the probability for the photon to exit port 1, and thus the conclusion that the gates anti-commute. The x-axis shows which gates have been implemented for $\hat{U}_A$-$\hat{U}_B$.
    }
    \label{fig:Commutation Game}
\end{figure}
In this section, we show additional measurements taken on this setup.

\subsection{Commutation / Anti-Commutation Game}
\label{sec::game}
Shortly after it was proposed that the quantum SWITCH offers a computational advantage over conventional quantum circuits for certain tasks \cite{ChiribellaSWITCH,Arajo2015}, one such task---the Commutation / Anti-Commutation Game---was experimentally realized with a two-party SWITCH \cite{Lorenzo}. This task takes the form of a game, in which a referee provides a player with two unitary gates $\hat{A}$ and $\hat{B}$. 
The referee states that these two unitaries gates either commute or anti-commute, and the player's goal is to decide which statement is true.
However, the player is only permitted to use each gate once.
When the player has access to a quantum SWITCH, they can win the game perfectly.
However, When the player is constrained to use a conventional (causally-ordered) quantum circuit, they would need to query one gate at least twice. 
The SWITCH result can be easily understood by looking at the output state of the SWITCH:
\begin{equation}
    \ket{\Psi} = \frac{1}{2}\ket{0}_C \acomm{\hat{A}}{\hat{B}} \ket{\Psi}_T + \frac{1}{2}\ket{1}_C\comm{\hat{A}}{\hat{B}} \ket{\Psi}_T \,.
\label{eq::unitary commutation state}
\end{equation}
Since the player is sure that either the commutator or anti-commutator is zero, by measuring the control and finding it in $\ket{0}_C$ they are sure the gates commute, while if they find it in $\ket{1}_C$ they can conclude that the gates anti-commute.

Before performing higher-order process tomography, we first verified the correct performance of the new architecture of our quantum SWITCH by implementing this game. To do so, Alice and Bob could set the Pauli operators $X=\sigma_X$, $Y=\sigma_Y$, $Z=\sigma_Z$ and the identity operator $I=\sigma_0$ with the well known commutation relation $\comm{\hat{\sigma}_i}{\hat{\sigma}_j} = 2i\epsilon_{ijk}\hat{\sigma}_k$ and anti-commutation relation $\acomm{\hat{\sigma}_i}{\hat{\sigma}_j} = 2\delta_ij\mathbb{1}_{2x2}$ \cite{NielsenChuang}. Fig.~\ref{fig:Commutation Game} shows the probabilities for measuring the control either in state $\ket{0}_C$ or $\ket{1}_C$.

The probabilities were calculating via measuring coincidence rates between the heralding detector and both output ports of the MZI with a coincidence window of $5$ ns between. Data was acquired for a total of $10$ s for each setting. The visibility of the interferometer was measured as $v^2 = 0.97$. We calculated the success probability for the photon leaving the correct port as
\begin{equation}
\begin{split}
	p_{\text{succ}} &= \frac{1}{2}(\bar{p}(0|[\cdot,\cdot]) + \bar{p}(1|\{\cdot,\cdot\})) \\
	& = 0.974 \pm 0.18 \,.
\end{split}
\end{equation}
Thus, our new architecture achieves a similar performance as previous implementations \cite{Lorenzo}, but now in a phase-stable manner.

\section{The definition of causal non-separability}\label{app:def_indefinite}

\begin{table*}[t]
\begin{center}
\begin{tabular}{|c || c | c | c | c |}
 \hline & & & & \\[-2ex]
    & ${G}_\text{y-,res}^\text{GR}$ & ${G}_\text{exp,res}^\text{GR}$ & ${G}_\text{y-,all}^\text{GR}$ & ${G}_\text{exp,all}^\text{GR}$  \\ [0.5ex] 
 \hline\hline
 $W^{y-}_s$ & $-0.500$ & $-0.483$ &$-0.500$ & $-0.484$\\ 
 \hline
 $W_\text{exp}$&$-0.346\pm0.003$&$-0.361\pm0.003$ & $-0.323\pm0.003$ & $-0.363\pm0.003$\\
[1ex] 
 \hline
\end{tabular}
\caption{\textbf{Generalized Robustness Alternative Definition Witness Analysis.} A summary of the different generalized robustness witnesses constructed using the definition of \cite{wei2019experimental}, presented in Eq.~\eqref{eq:wechs}.  The witnesses $G_{i,j}$ are labelled by two subscripts. The first indicates if the witness was designed for the ideal process matrix~$W^{y-}$ (subscript ''$y-$'') or the experimental process matrix $W_\text{exp}$ (subscript ''exp''). The second subscript indicates if the restricted measurement set (subscript ''res'') or the complete measurement set (subscript ''all'') was used for the witness. 
The first row shows the value of the witness for $W^{y-}$, and the second row shows the experimental values, which were evaluated as $\Tr[G_{i,j} W_\text{exp}]$. }
\label{tab:GR2}
\end{center}
\end{table*}

\begin{table*}[t]
\begin{center}
\begin{tabular}{|c || c | c | c | c |} 
 \hline & & & & \\[-2ex]
    & ${G}_\text{y-,res}^\text{WN}$ & ${G}_\text{exp,res}^\text{WN}$ & ${G}_\text{y-,all}^\text{WN}$ & ${G}_\text{exp,all}^\text{WN}$  \\ [0.5ex] 
 \hline\hline
 $W^{y-}_s$ & $-0.828$ & $-0.800$ &$-1.000$ & $-0.965$\\ 
 \hline
 $W_\text{exp}$&$-0.572\pm0.005$&$-0.614\pm0.006$ & $-0.689\pm0.007$ & $-0.739\pm0.008$\\
[1ex] 
 \hline
\end{tabular}
\caption{\textbf{White Noise Alternative Definition Witness Analysis.} A summary of the different white noise witnesses constructed using the definition of \cite{wei2019experimental}, presented in Eq.~\eqref{eq:wechs}.  The witnesses $G_{i,j}$ and process matrix labels are labelled by two subscripts described in the caption of Tab. \ref{tab:GR2}.}
\label{tab:WN2}
\end{center}
\end{table*}

Let $W\in~\L\left(\H_P \otimes \H_{A_\text{in}}\otimes\H_{A_\text{out}}\otimes\H_{B_\text{in}}\otimes\H_{B_\text{out}}\otimes\H_{F}\right)$ be a bipartite process matrix which a common past and common future. A process $W^{A>B}$ is said to be causally ordered from $A$ to $B$ if it can be constructed by beans of a sequential quantum circuit~\cite{oreshkov2012quantum,Arajo2015}, also called a quantum comb~\cite{chiribella08_arquitechture,Chiribella2013}). More explicitly, a process matrix $W^{A>B}\in~\L\left(\H_P \otimes \H_{A_\text{in}}\otimes\H_{A_\text{out}}\otimes\H_{B_\text{in}}\otimes\H_{B_\text{out}}\otimes\H_{F}\right)$ is causally ordered from $A$ to $B$ if it respects
\begin{align}
\tr_{F}(W^{A>B})=& \tr_{B_O F}(W^{A>B})\otimes\frac{\id_{B_O}}{d_{B_O}} \\
\tr_{B_IB_OF}(W^{A>B})=& \tr_{A_OB_IB_OF}(W^{A>B})\otimes\frac{\id_{A_O}}{d_{A_O}} \\
\tr_{A_IA_OB_IB_OF}(W^{A>B})=& \tr_{PA_IA_OB_IB_OF}(W^{A>B})\otimes\frac{\id_{P}}{d_{p}}.
\end{align}

Analogously, a process matrix $W^{B>A}\in~\L\left(\H_P \otimes \H_{A_\text{in}}\otimes\H_{A_\text{out}}\otimes\H_{B_\text{in}}\otimes\H_{B_\text{out}}\otimes\H_{F}\right)$ is causally ordered from $B$ to $A$ if it respects
\begin{align}
\tr_{F}(W^{B>A})=& \tr_{A_O F}(W^{B>A})\otimes\frac{\id_{A_O}}{d_{A_O}} \\
\tr_{A_IA_OF}(W^{B>A})=& \tr_{B_OA_IA_OF}(W^{B>A})\otimes\frac{\id_{B_O}}{d_{B_O}} \\
\tr_{A_IA_OB_IB_OF}(W^{B>A})=& \tr_{PA_IA_OB_IB_OF}(W^{B>A})\otimes\frac{\id_{P}}{d_{p}}.
\end{align}

In the main part of this work, we follow the definition of Ref.~\cite{rubino2017}, where a bipartite process matrix $W\in~\L\left(\H_P \otimes \H_{A_\text{in}}\otimes\H_{A_\text{out}}\otimes\H_{B_\text{in}}\otimes\H_{B_\text{out}}\otimes\H_{F}\right)$  with common past and common future is said to be causally separable if it can be written as a convex sum of causally-ordered process matrices. That is, if we can write
\begin{align} \label{eq:def1app}
    W = p W^{A>B} + (1-p) W^{B>A},
\end{align}
where $p\in[0,1]$, and $W^{A>B}$ and $W^{B>A}$ are causally-ordered process. However, as discussed earlier, there exists a non-equivalent definition, which go beyond simple convex combination and allow incoherent classical control of causal orders. This definition is presented in Ref.~\cite{wechs2019_definition}, and it is proven to be equivalent to the notion of extensible causal, presented in Ref.~\cite{giarmatzi15_causallySEP}. In the bipartite scenario with a common past and common future, Ref.~\cite{wechs2019_definition} states that a process matrix $W\in~\L\left(\H_P \otimes \H_{A_\text{in}}\otimes\H_{A_\text{out}}\otimes\H_{B_\text{in}}\otimes\H_{B_\text{out}}\otimes\H_{F}\right)$ is causally separable when there exists causally ordered processes $W^{A>B}$ and $W^{B>A}$ such that,
\begin{align} \label{eq:wechs}
  \frac{\id_P}{d_P} \otimes  \tr_{P}(W)= p W^{A>B} + (1-p) W^{B>A}.
\end{align}
Notice that the definition of Ref.~\cite{wechs2019_definition}  presented in Eq.~\ref{eq:wechs} is more relaxed than the one of Ref.~\cite{rubino2017}, and presented in \eqref{eq:def1app}. Indeed, as we show latter, there are processes which are causally separable following the definition of \cite{wechs2019_definition}, but are causally non-separable following the definition of \cite{rubino2017}.

In the definition of Ref.~\cite{wechs2019_definition}, in order for a causal witness to be valid, one should include an extra constraint, which reads as
\begin{align}
    G=\frac{\id_P}{d_P}  \otimes \tr_P(G).
\end{align}
We can then re-calculate the numbers presented in Table~\ref{tab:GR} and Table~\ref{tab:WN} from the main text, but following the definition of Ref.~\cite{wechs2019_definition}. These results are presented in Table~\ref{tab:GR2} and Table~\ref{tab:WN2}, and we notice that although there are some difference in the obtained numbers, the qualitative result remains the same.
\end{document}